\documentclass{aa}  
\usepackage[varg]{txfonts}
\usepackage{microtype}
\linenumbers 

\usepackage{natbib}
\bibpunct{(}{)}{;}{a}{}{,} 

\usepackage{graphicx}
\usepackage{txfonts}
\usepackage{tikz}

\usepackage{units}
\usepackage[breaklinks=true]{hyperref}
\usepackage[normalem]{ulem}

\makeatletter 
\renewcommand*\aa@pageof{, page \thepage{} of \pageref*{LastPage}}
\makeatother

\begin{document}

   \title{Tidal disruption events as the origin of\\the \textit{eROSITA} and \textit{Fermi} bubbles}


   \author{Tassilo Scheffler \inst{\ref{inst1},\ref{inst2}}\href{https://orcid.org/0009-0002-3600-4516}{\includegraphics[width=9pt]{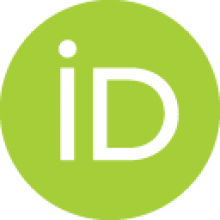}}
    \and
    Michael M.~Schulreich\inst{\ref{inst1}}\href{https://orcid.org/0000-0001-7761-9766}{\includegraphics[width=9pt]{Orcid-ID.png}}
    \and
    David P.~P.~R.~Schurer\inst{\ref{inst1}}
    \and
    Dieter Breitschwerdt\inst{\ref{inst1}}\href{https://orcid.org/0000-0001-6718-8069}{\includegraphics[width=9pt]{Orcid-ID.png}}
    }

   \institute{Zentrum für Astronomie und Astrophysik, Technische Universität Berlin, Hardenbergstraße 36, D-10623 Berlin, Germany\label{inst1}\\
        \and
        Sterrenkundig Observatorium, Ghent University, Krĳgslaan 281-S9, B-9000 Ghent, Belgium\label{inst2}\\
        \email{\href{mailto:tassilo.scheffler@ugent.be}{tassilo.scheffler@ugent.be}}
    }

   \date{Received 23 August 2024 / Accepted 29 January 2025}

 
  \abstract{The recently discovered spherical \textit{eROSITA} bubbles extend up to a latitude of $\pm80^\circ$--$85^\circ$ in the X-ray regime of the Milky Way halo. Similar to the $\gammaup$-ray \textit{Fermi} bubbles, they evolve around the Galactic center, making a common origin plausible. However, the driving mechanism and evolution of both bubbles are still under debate.}
  {We investigate whether hydrodynamic energy injections at the Galactic center, such as tidal disruption events, could have inflated both bubbles. The supermassive black hole Sagittarius A$^*$ is expected to tidally disrupt a star every 10--\unit[100]{kyr}, potentially leading to an outflow from the central region that drives a shock propagating into the Galactic halo due to its vertically declining density distribution, ultimately forming a superbubble that extends out of the disk similar to the \textit{eROSITA} and \textit{Fermi} bubbles.}
  {We model tidal disruption events in the Galaxy using three-dimensional hydrodynamical simulations, considering different Milky Way mass models and tidal disruption event rates. We then generate synthetic X-ray maps and compare them with observations.}
  {Our simulation results of a $\betaup$-model Milky Way halo show that superbubbles, blown for \unit[16]{Myr} by regular energy injections at the Galactic center that occur every \unit[100]{kyr}, can have a shape, shell stability, size, and evolution time similar to estimates for the \textit{eROSITA} bubbles, and an overall structure reminiscent of the \textit{Fermi} bubbles. The $\gammaup$-rays in our model would stem from cosmic ray interactions at the contact discontinuity, where they were previously accelerated by first-order Fermi acceleration at in situ shocks.}
  {Regular tidal disruption events in the past 10--20 million years near the Galactic center could have driven an outflow resulting in both, the X-ray emission of the \textit{eROSITA} bubbles and the $\gammaup$-ray emission of the \textit{Fermi} bubbles.}


  \authorrunning{Tassilo Scheffler et al.}
  \maketitle
%
\section{Introduction}

  Almost one and a half decades ago, the all-sky \textit{Fermi} Large Area Telescope discovered the $\gammaup$-ray \textit{Fermi} bubbles (FBs) in the Milky Way halo in the GeV energy range \citep{Su2010}. More recently, the larger X-ray \textit{eROSITA} bubbles (EBs) were discovered in the 0.6--\unit[1]{keV} regime \citep{Predehl2020}, also stretching into the halo \citep{Liu2024}. As both arise from the Galactic center, it is theorized that they have the same origin \citep[see e.g.][]{Sarkar2024}. The exact driving mechanism that generated a total thermal energy of about \unit[2.6$\times$10$^{56}$]{erg} \citep{Predehl2020} is, however, still under debate. 
  
  The X-ray emission can be explained by bremsstrahlung from rapidly decelerated particles in the shock front of the superbubbles \citep{Yang2022}. The $\gammaup$-ray emission of the FBs is non-thermal. Cosmic rays (CRs), accelerated in the halo, are necessary to produce the high-energy photons \citep{Su2010}. Different phenomena can provide enough energy for CR acceleration. An active galactic nucleus (AGN) jet, for example, can accelerate CR electrons in the Galactic center and transport them rapidly into the halo where they scatter with the interstellar radiation field by inverse Compton emission \citep{Yang2022}. Regular starbursts or AGN winds can also deliver enough energy for the acceleration of CRs in the Galactic center according to the hadronic wind model. Since CR leptons lose their energy faster than hadrons, only CR protons can reach the halo where they would get scattered by interstellar matter, producing neutral pions ($\piup^0$) that decay into $\gammaup$-rays, unless CR electron reacceleration in the Galactic halo occurs. The third possibility for CR acceleration in the Galactic halo is the in situ acceleration. Here, like in diffusive shock acceleration, CRs drifting through a magnetized plasma can resonantly generate magnetohydrodynamic waves via a so-called streaming instability, which leads to a strong scattering and a secular gain of CR energy after multiple reflections across the converging flow of a shock wave \citep{Axford1977,Bell1978,Blandford1978}. Another kind of in situ acceleration can be provided by the shock drift mechanism, resulting from gradient drift across the shock and to an $\vec{E} \times \vec{B}$-electric field component, which is strongest for perpendicular shocks, to accelerate particles at a high rate \citep{Jokipii1982}.

  Hadronic models appear unlikely because both their spectrum of the $\piup^0$ decayed $\gammaup$-rays and their synchrotron radiation spectrum of the secondary leptons are too soft \citep{Su2010,Ackermann2014}. High altitude water Cherenkov (\textit{HAWC}) observations of $\gammaup$-rays in the northern FBs also disfavor hadronic models since their flux of high energy $\gammaup$-rays is too low \citep{Abeysekara2017}.

  Many numerical studies are able to reproduce the $\gammaup$-ray FBs \citep[e.g.][]{Yang2012, Guo2012, Crocker2015, Sarkar2019, Ko2020, Zhang2020}. The X-ray EBs, however, have a different emission energy, shape, size, and contain ten times more thermal energy than the FBs, making it necessary to significantly adjust all FB models. New approaches must also be able to explain the two different bubbles simultaneously. Due to these challenges, to our knowledge only \citet{Yang2022}, \citet{Chang2024}, and \citet{Tseng2024} were able to numerically reproduce both the EBs and FBs simultaneously.
  
  All of these studies assume that the supermassive black hole (SMBH) Sagittarius A$^*$ (Sgr~A$^*$) in the Galactic center underwent an active phase with near-Eddington luminosity in which it ejected relativistic jets into the Milky Way halo. The AGN jets within the models start a few million years ago and last from several hundred thousands to a few million years. After the active phase of the jets, they expand into an X-ray emitting spherical cocoon shape with $\gammaup$-ray emitting CRs in the interior, thus resembling the EBs and FBs simultaneously.
  
  Although AGN jet models are able to explain both the EBs and FBs at the same time, the present weakness of the radio source at Sgr~A$^*$ does not indicate an active phase of the central SMBH only a few million years ago. Furthermore, most AGN jet models assume that the jets are orthogonal to the Milky Way disk. Thus, the accretion disk of Sgr~A$^*$ would have needed to be co-planar with the Milky Way disk, since AGN jets are ejected perpendicular to the accretion disk of their SMBH. However, the \citet{EHT2022} discovered that the disk of Sgr~A$^*$ has an inclination to the line of sight from the Earth and hence to the Milky Way disk. Therefore, relativistic jets of Sgr~A$^*$ should have been tilted to the Galactic disk. Numerical investigations by \citet{Sarkar2023} show that tilted jets can produce spherical bubbles only if the jets are strongly dissipated. \citet{Dutta2024} use numerical models of clumpy media to rule out dissipation in the clumpy Milky Way halo, meaning that the dissipation had to occur in the central molecular zone (CMZ) or by magnetically dominated jets \citep[see also][and references therein]{Sarkar2024}. \citet{Tseng2024} can reproduce the EBs and FBs with tilted, CMZ dissipated jets, however, \citet{Dutta2024} argue that the imprint of recently dissipated jets would be visible in the CMZ today and that magnetic dissipation should have resulted in a stronger forward shock than observed in the EBs. Furthermore, \citet{Mou2023} find that tilted AGN jets cannot reproduce the asymmetric brightness distribution of the EBs. Thus, future studies need to explore in greater detail whether the EBs and FBs could still originate from dissipated jets. 

  Stellar feedback, especially from supernovae (SNe) could also reach luminosities high enough to lead to a thermal outflow, as seen in the M82 galaxy \citep{Rieke1980}. The CMZ in the Milky Way, however, is assumed to have a much lower nuclear SN rate of $0.02$--$0.15$ per century \citep{Crocker2011,Ponti2015}. With an average SN energy of $\unit[10^{51}]{erg}$, a maximum luminosity of $\unit[4.8\times 10^{40}]{erg\,s^{-1}}$ can be achieved. This luminosity could blow the FBs in a few tens of millions of years, which is why different models consider star formation as their origin \citep[e.g.][]{Crocker2015,Sarkar2019}. The total thermal energy of the EBs, however, would be reached after a minimum of about \unit[170]{Myr}. This time scale is significantly longer than the predicted current age of the EBs that range from a few Myr to a maximum of \unit[20]{Myr} \citep{Predehl2020,Schulreich2022,Yang2022,Mou2023}. Similar to \citet{Mou2023}, who argue that SNe in the CMZ lose too much energy to drive a thermal outflow due to radiative cooling, we conclude that SNe alone are most likely not the origin of the EBs.

  On a different note, SMBHs such as Sgr~A$^*$ are assumed to tidally disrupt nearby stars on a regular basis due to their high differential gravitational forces \citep{Lacy1982, Cheng2011}. Only about half of the debris is accreted to the SMBH, so that tidal disruption events (TDEs) typically release large amounts of energy, scaling with the stellar and SMBH masses \citep{Alexander2005}. In galaxies similar to the Milky Way, TDEs are expected to occur every 10--\unit[100]{kyr} \citep{Magorrian1999}. Numerical simulations by \citet{Ko2020} show that the energy of regular TDEs is sufficient to blow superbubbles into the Milky Way halo, assuming a luminosity of \unit[3$\times$10$^{41}$]{erg\,s$^{-1}$}. However, they only studied the FBs in simplified Milky Way halo models and did not consider the larger EBs.

  In this paper, we perform numerical studies of the TDE model in which $\gammaup$-rays are accelerated in situ, and X-rays result from a strong forward propagating shock, strengthening in a halo with decreasing density, thereby compressing and heating the gas. In Sect.~\ref{sec:methods}, we describe our numerical setup. Section \ref{sec:results} shows the simulation results and compares the derived synthetic X-ray maps with observations. We summarize the main results in Sect.~\ref{sec:conclusions}.

\section{Methods} \label{sec:methods}
  Hydrodynamical simulations of the TDE model are performed using the publicly available adaptive mesh refinement (AMR) code \texttt{RAMSES}\footnote{\url{https://bitbucket.org/rteyssie/ramses}} \citep{Teyssier2002}. The code uses the second-order accurate monotonic upstream-centered scheme for conservation laws (MUSCL), which is an extension of Godunov's method \citep{Toro1997} to solve the Euler equations with the aid of an approximate Harten-Lax-van Leer-contact (HLLC) Riemann solver \citep{Toro1992,Toro1994}. To model the evolution of the superbubbles as realistically as possible, three-dimensional simulations are used. The simulation box is chosen large enough so that the bubbles never reach the boundaries during their evolution.

  In a resolution study, we investigated the necessary refinement of the AMR grid to reduce numerical diffusion \citep{Patankar1980}. The refinement level of each cell was determined by the pressure and density gradient to its neighboring cells to follow the shocks and contact discontinuities in great detail. We found that the shape of the superbubbles converges at a maximum refinement level of $\sim$\unit[9.8]{pc}. The minimum resolution in the AMR grid was set to \unit[312.5]{pc}.

  \begin{figure*}[hbpt!]
     \centering
     \begin{minipage}[t]{0.33\linewidth}
     \includegraphics[width=\linewidth]{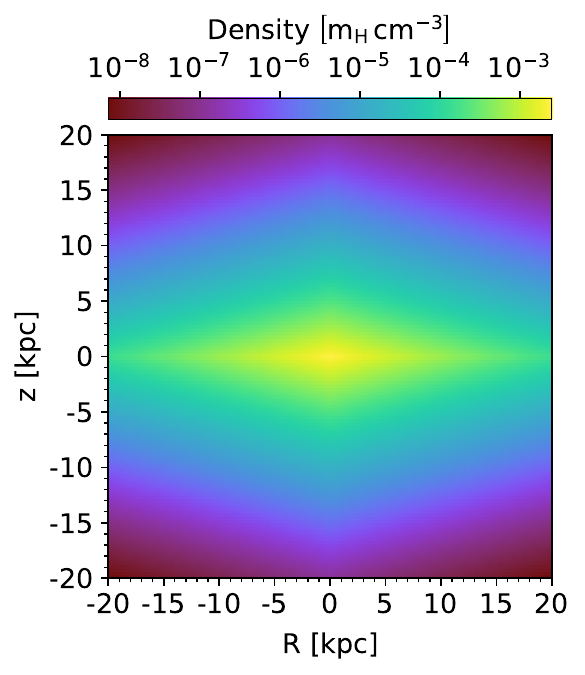}
     \end{minipage}
     \begin{minipage}[t]{0.33\linewidth}
     \includegraphics[width=\linewidth]{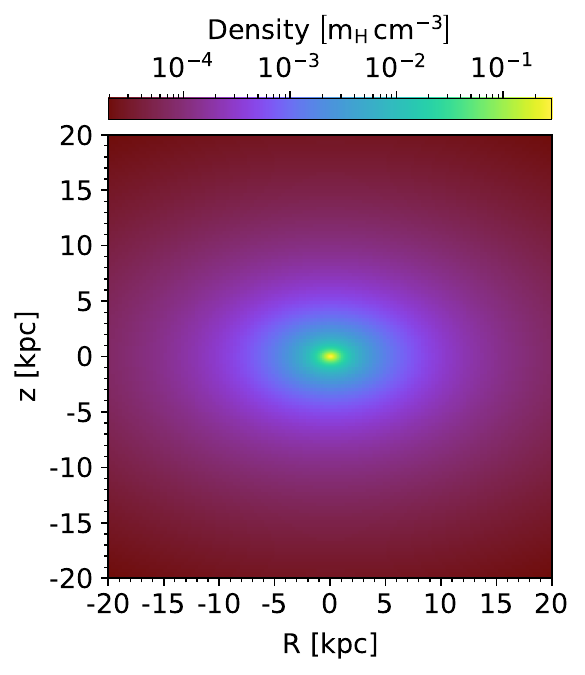}
     \end{minipage}
     \begin{minipage}[t]{0.33\linewidth}
     \includegraphics[width=\linewidth]{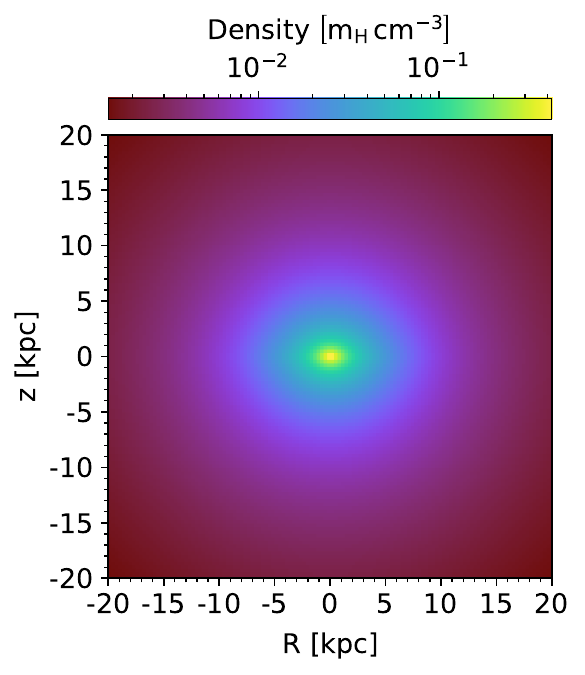}
     \end{minipage}
     
      \caption{Density distributions of the different Milky Way models dependent on the height $z$ and the radius $R$ away from the Galactic center. Two-dimensional slices through the origins of the three-dimensional simulation boxes are shown. The exponential halo model (left), the $\betaup$-model halo (middle), and the multicomponent model (right) are described by Eqs.~\ref{eq:exp_density_profile_full}, \ref{eq:beta_density_profile_full}, and \ref{eq:dens_McMillan1}--\ref{eq:dens_McMillan4}, respectively. The models differ by several orders of magnitude in their density distribution.}
      \label{fig:halos}
  \end{figure*}

  The energy and mass outflows from TDEs occur on a much smaller scale than the resolution of our simulations since stars are tidally disrupted within the Roche radius of an SMBH, which is smaller than a parsec for the SMBH at the Galactic center \citep{Lacy1982}. Thus, we modeled TDEs as simple energy and mass injections that are homogeneously distributed in the central eight grid cells. The edge length of the cubic injection region (in which the gas speed was set to be zero) was therefore $\sim$\unit[20]{pc}, whereas the entire cubic computational domain had an extension of \unit[40]{kpc}. Given that TDEs in the Milky Way occur every 10--\unit[100]{kyr} on average, we injected energy and mass either every 10 or \unit[100]{kyr}, to probe the two limiting cases. The energy of a single TDE, either 9.5$\times$10$^{52}$ or \unit[9.5$\times$10$^{53}$]{erg} in our model, was injected in a single time step, resulting in the same average luminosity of \unit[3$\times$10$^{41}$]{erg\,s$^{-1}$} in both cases. The chosen luminosity was based on energy estimates for the EBs \citep{Predehl2020} and TDE events in the Milky Way \citep{Ko2020}. During our numerical investigations, we found that the injected mass hardly influences the shape of the bubbles but was necessary for preventing the interior density, and thus, due to our refinement criterion, also the simulation time step, from approaching progressively smaller values. We chose to set the injected mass to the respective initial density in the central cells of the selected Milky Way model.

  The gas in our simulation obeys the ideal gas law. It is initially in an isothermal state. We do not consider self-gravity, but instead assume a hydrostatic equilibrium at the beginning of our simulations that is established by an external gravitational field counteracting the thermal pressure gradient. This assumption is necessary to keep the background halo at rest during the simulations. Gravitation as an external force field does not directly affect the dynamics of the bubbles, since its potential energy is much smaller than the energies of the TDEs and superbubbles, but it shapes the density distribution in the halo via the hydrostatic equilibrium, which indirectly feeds back on the bubble dynamics. The density distribution $\rho(R,z)$, dependent on the height $z$ above the Galactic disk and the distance $R$ in radial disk direction, is then related to the gravitational potential $\Phi(R,z)$ by
  \begin{equation}      
  \rho(R,z)=\rho_0\,\exp{\left(\frac{\Phi_0-\Phi(R,z)}{k_\text{B}\,T/(\mu\,m_\text{H})}\right)}\,,\label{eq:mc_model}
  \end{equation}
  with $\rho_0$, $\Phi_0$, $k_\text{B}$, $T$, and $\mu=0.62$ being the the central density and gravitational potential, Boltzmann's constant, the temperature, and the mean molecular weight for an ionized gas with solar abundances in units of the hydrogen atomic mass $m_\text{H}$, respectively.
  
  The background density distribution was found to have a strong influence on the structure of the evolving superbubbles. Therefore, different axisymmetric Milky Way mass models were considered.
  
  First, we used a simple exponential halo model \citep[e.g.][]{Cordes1991,Nakashima2018} with a distribution of
  \begin{equation}
      \rho(R,z)=n_0\,\mu\,m_\text{H}\,\exp{\left(-\frac{R}{R_0}\right)}\,\exp{\left(-\frac{z}{z_0}\right)}\,.\label{eq:exp_density_profile_full}
  \end{equation}
  Its parameters include the central number density $n_0=\unit[0.004]{cm^{-3}}$, the scale radius $R_0=\unit[7]{kpc}$, and the scale height $z_0=\unit[2]{kpc}$ \citep{Nakashima2018}. The temperature chosen for this halo model was $T=\unit[10^6]{K}$. Equation \ref{eq:exp_density_profile_full} is plotted in the left panel of Fig.~\ref{fig:halos}.
  
  Secondly, the $\betaup$-model halo proposed by \citet{Miller2013} was considered. It is based on \ion{O}{VII} and \ion{O}{VIII} absorption line observations and assumes a modified King profile, which shows good agreement with the density distribution of other galaxies \citep{Osullivan2003}. It is given by 
  \begin{equation}
      \rho(R,z)=n_0 \,\mu\,m_\text{H}\,\left[1+\left(\frac{R}{R_0}\right)^2+\left(\frac{z}{z_0}\right)^2\right]^{-3\,\beta/2}\,,\label{eq:beta_density_profile_full}
  \end{equation}
  with $n_0=\unit[0.46]{cm^{-3}}$, $R_0=\unit[420]{pc}$, $z_0=\unit[260]{pc}$, and a galaxy-dependent factor, constrained to $\beta=0.71$ for the Milky Way. The corresponding plot is shown in the central panel of Fig.~\ref{fig:halos}. As for the exponential model, the temperature was set to $T=\unit[10^6]{K}$.

  Finally, the multicomponent model of \citet{McMillan2017} was investigated, according to which the total gravitational potential of the Milky Way can be assembled by summing up six individual components. Only the density distribution obtained via the Poisson equation is shown here since the corresponding gravitational potential cannot be described analytically. The density distribution of the hydrostatic equilibrium needed to be calculated from the numerically given gravitational potential via Eq.~\ref{eq:mc_model} with $\rho_0=\unit[0.03]{cm^{-3}}\,\mu\,m_\text{H}$ \citep{Zhang2020} and $T=\unit[2.32\times10^{6}]{K}$ (corresponding to a thermal energy of \unit[0.2]{keV}). The first two components are the thin and thick stellar disks that are described by 
  \begin{equation}
      \rho_\text{d}(R,z)= \frac{\Sigma_0}{2z_\text{d}}\exp\left(-\frac{|z|}{z_\text{d}}-\frac{R}{R_\text{d}}\right)\,,\label{eq:dens_McMillan1}
  \end{equation}
  where $\Sigma_{0}=\unit[896]{M_\odot\,pc^{-2}}$ is the central surface density, $z_\text{d}=\unit[300]{pc}$ is the scale height, and $R_\text{d}=\unit[2.5]{kpc}$ is the scale length for the thin disk, and $\Sigma_{0}=\unit[183]{M_\odot\,pc^{-2}}$, $z_\text{d}=\unit[900]{pc}$, and $R_\text{d}=\unit[3.02]{kpc}$ for the thick disk. The next components are the \ion{H}{I} and H$_2$ gas disks which follow 
  \begin{equation}
      \rho_\text{g}(R,z)=\frac{\Sigma_0}{4\,z_\text{g}}\,\exp\left(\frac{R_\text{m}}{R}-\frac{R}{R_\text{g}}\right)\,\text{sech}^2\left(\frac{z}{2\,z_\text{g}}\right)\,,
  \end{equation}
  with $\Sigma_{0}=\unit[53.1]{M_\odot\,pc^{-2}}$, $z_\text{g}=\unit[85]{pc}$, $R_\text{g}=\unit[7]{kpc}$, and $R_\text{m}=\unit[4]{kpc}$ for the \ion{H}{I} disk, and $\Sigma_{0}=\unit[2180]{M_\odot\,pc^{-2}}$, $z_\text{g}=\unit[45]{pc}$, $R_\text{g}=\unit[1.5]{kpc}$, and $R_\text{m}=\unit[12]{kpc}$ for the H$_2$ disk, respectively. The bulge of the Milky Way was modeled as 
  \begin{equation}
      \rho_\text{b}(R,z) = \frac{\rho_{0,\text{b}}}{(1+\sqrt{R^2+4\,z^2}/r_0)^\alpha}\exp{\left[-\left(\frac{\sqrt{R^2+4\,z^2}}{r_\text{cut}}\right)^2\right]}\,,\label{eq:dens_McMillan3}
  \end{equation}
  with $\rho_{0,\text{b}}=$ \unit[9.93$\times$10$^{10}$]{M$_\odot$\,kpc$^{-3}$}, $r_0=\unit[0.075]{kpc}$, $\alpha=1.8$, and $r_\text{cut}=\unit[2.1]{kpc}$. The last component is the dark matter halo which is described by 
  \begin{equation}
      \rho_\text{h}(R,z)=\frac{\rho_{0,\text{h}}}{(r/r_\text{h})(1+r/r_\text{h})^2}\,,\label{eq:dens_McMillan4}
  \end{equation}
  where $r=\sqrt{R^2+z^2}$, $\rho_{0,\text{h}}=\unit[0.00854]{M_\odot\,pc^{-3}}$, and $r_\text{h}=\unit[19.6]{kpc}$. The density distribution for this model is shown in the right panel of Fig.~\ref{fig:halos}.
  
  All three Milky Way density models were studied in separate numerical simulations under the same assumptions. To keep the computational effort low, radiative cooling is not included in our model. It would lead to a thinner and denser shell and would have an impact on local structures but would not strongly influence the global shape of the superbubbles. Furthermore, magnetic fields are neglected. An initially disk-parallel magnetic field would wrap around the bubbles so that it would be tangential to the contact discontinuity, thereby strongly inhibiting Rayleigh-Taylor (RT) instabilities \citep{Breitschwerdt2000}. Furthermore, magnetic fields would decelerate the expansion due to magnetic tension forces. Our simulations should therefore be taken as a lower limit with respect to the dynamical time scale. Significant changes to the shape of the bubbles are not expected, since the energy input by TDEs vastly exceeds the magnetic energy stored in the disk field. As bubble expansion progresses, and the ambient density decreases (see above), the frozen-in magnetic field diminishes accordingly.

  From the most promising simulation results, synthetic \textit{eROSITA} maps were calculated, showing the expected X-ray emission of the simulated superbubbles. A collisional ionization equilibrium source model for X-ray emission of the intergalactic medium (IGM) is considered, using the \texttt{pyXSIM}\footnote{\url{https://hea-www.cfa.harvard.edu/~jzuhone/pyxsim/}} implementation \citep{ZuHone2014} of the \texttt{Cloudy} algorithm \citep{Chatzikos2023}. For the cooler, less relevant regimes outside of the main shock, a different IGM model including photoionization is used \citep{Khabibullin2019}. Absorption was simulated using the Tuebingen-Boulder ISM absorption model \citep{Wilms2000}, a variable foreground column density $N_\text{H}$, based on data from the \citet{HI4PI2016}, and the abundance table by \citet{Feldman1992} using a constant solar metallicity. Different metallicities do not change the general X-ray bubble structure but shift the intensity distribution slightly. With a lower metallicity, which is expected in the Milky Way halo, especially the background emission becomes more significant. Of the initially generated X-ray photons that reach the internal observer positioned in the Solar System, only a random sample was projected into an all-sky map using \texttt{pyXSIM}. Then, the instrumental effects were replicated with the Simulated Observations of X-ray Sources (\texttt{SOXS})\footnote{\url{https://hea-www.cfa.harvard.edu/soxs/}} \citep{ZuHone2023} software package, using the response matrix files, ancillary response files, point-spread functions, and particle background file of \textit{eROSITA}, distributed with the Simulation of X-ray Telescopes (\texttt{SIXTE})\footnote{\url{https://www.sternwarte.uni-erlangen.de/sixte/}} software 
  \citep{Dauser2019}. We also assumed that the filter of every \textit{eROSITA} camera was active, as this is the standard observing mode \citep{Predehl2021}. For the exposure time within the all-sky survey, we used a constant value, large enough to see the significant features in the final maps, because the reproduction of the varying exposure time of \textit{eROSITA} \citep{Predehl2021} is beyond the scope of this paper. Lastly, only photons with energies between \unit[0.6]{keV} and \unit[1]{keV} were selected for the synthetic X-ray maps using the \texttt{Python} implementation \texttt{healpy} \citep{Zonca2019} of \texttt{HEALPix}\footnote{\url{http://healpix.sourceforge.net}} \citep{Gorski2005}.


\section{Results} \label{sec:results}

  To start our analysis, we first performed simulations in a simple homogeneous ambient medium and compared it to the corresponding analytical solution for a wind-blown bubble \citep{Castor1975,Weaver1977}. Our numerical results were in very good agreement with the theory. Next, we included a simplified exponential and a simplified $\betaup$-model halo in our simulation, which can be analytically solved using the Kompaneets formalism \citep{Kompaneets1960,Baumgartner2013,Ko2020,Schulreich2022}. Although the ellipsoidal structure of the exponential halo and the more spherical structure of the $\betaup$-model halo bubbles were successfully reproduced in these simulations, the numerically resulting bubbles were smaller than the theoretical predictions at any given time step. This deviation occurred because we included fewer assumptions in our model than \citet{Kompaneets1960}. For example, we accounted for the surrounding pressure, which led to a noticeable deceleration and resulted in smaller bubble sizes in our model. Therefore, we can conclude that our numerical model reproduced the analytically given features. Hence, more complex and realistic Milky Way models could be implemented into the model.

\subsection{Exponential halo model}

  In the exponential halo model, mass and energy injections were initially set to happen at each time step instead of every 10 or \unit[100]{kyr} to study the general shape of the bubble, which was not significantly altered by changing the TDE rate with similar luminosity. The time step varied so that energy and mass were injected every 100--\unit[1000]{yr}. The simulation was stopped at \unit[12.5]{Myr} because the approximate present-day size of the EBs was reached. In total, an energy of \unit[1.2$\times$10$^{56}$]{erg} was injected. 
   
\begin{figure*}[hbtp!]
      \centering
      \begin{minipage}[t]{0.48\linewidth}
      \includegraphics[width=\linewidth]{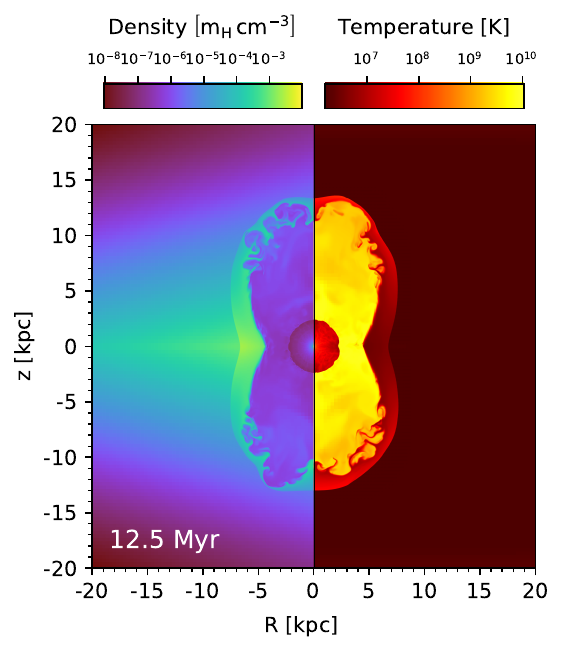}
      \end{minipage}
      \begin{minipage}[t]{0.48\linewidth}
      \includegraphics[width=\linewidth]{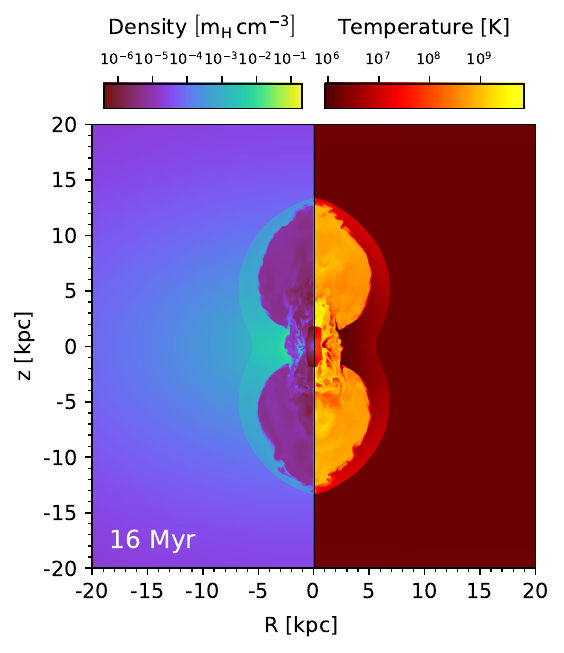}
      \end{minipage}
      \begin{minipage}[t]{0.48\linewidth}
      \includegraphics[width=\linewidth]{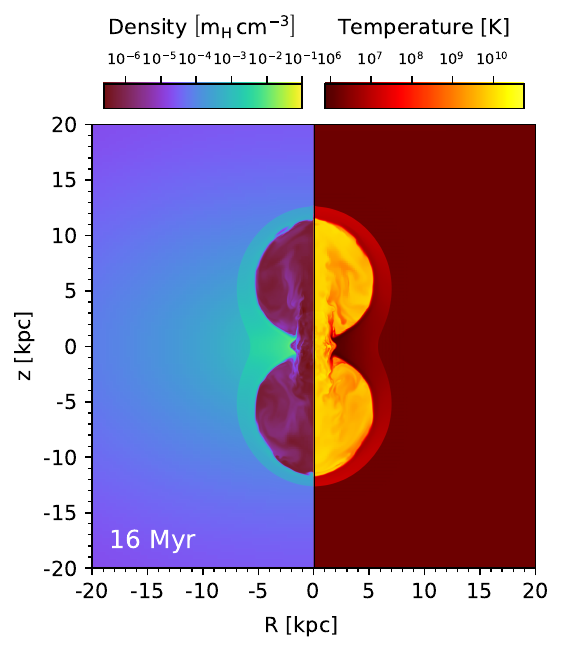}
      \end{minipage}
      \begin{minipage}[t]{0.48\linewidth}
      \includegraphics[width=\linewidth]{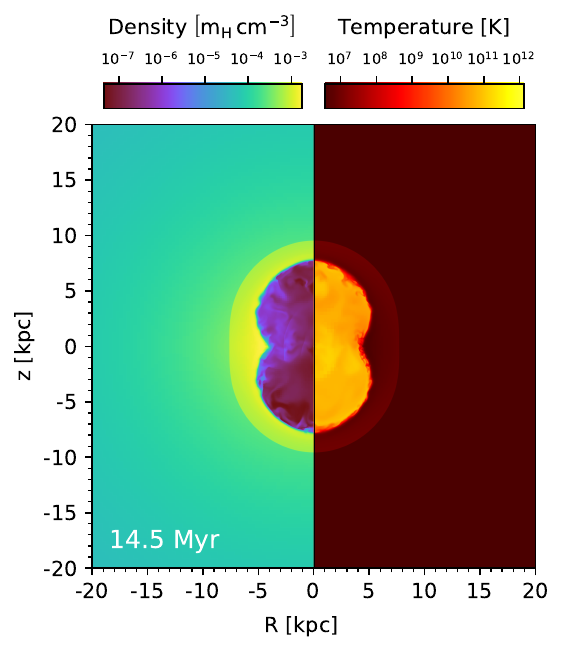}
      \end{minipage}
      \caption{Resulting density distributions of the numerical simulations for different Milky Way models and TDE rates. The snapshots are taken at the end of the simulations when the forward shocks reach the size of the EBs and show the detailed bubble structure, including the elapsed time since the first TDE. The different subfigures correspond to the results of the exponential halo model with energy and mass injections every 100--\unit[1000]{yr} (top left), the $\betaup$-model halo with TDEs every \unit[10]{kyr} (top right), the $\betaup$-model halo with TDEs every \unit[100]{kyr} (bottom left), and the multicomponent model with TDEs every \unit[100]{kyr} (bottom right).}
      \label{fig:results}
  \end{figure*}

  The result can be seen in the top-left panel of Fig.~\ref{fig:results}. Although the forward shock agrees well with the spherical shape of the EBs, the shell and especially the contact discontinuity of the superbubbles are strongly eroded by fast growing RT instabilities, which arise because the density gradient of the exponential halo is opposite to the shock acceleration \citep{Rayleigh1882,Taylor1950}. 
  
  Following the in situ model for CR acceleration, the contact discontinuity should resemble the FBs in our model. After the CR protons and electrons gain energy through first-order Fermi acceleration at the strong termination shock ($\mathcal{M}>20$, with $\mathcal{M}=v/c$, where $v$ is the local flow speed and $c$ the local sound speed) visible in the inner 3--\unit[5]{pc}, and at individual shocks propagating into the halo due to consecutive TDEs, they can penetrate the low-density interior of the bubbles. The $\gammaup$-ray emission could then come from inverse Compton scattering of high energy electrons on the cosmic microwave background (or other low-energy) photons, synchrotron emission from electrons and positrons in strong magnetic fields, CR proton interactions with low-energy protons leading to $\piup^0$-decay, and bremsstrahlung.

  Through the strong erosion of the contact discontinuity in the exponential halo model bubble, this $\gammaup$-ray emission would certainly indicate instabilities in observations even if small perturbations were smoothed out. However, \textit{Fermi} suggests that the FBs do not incorporate large instabilities. Similarly, \textit{eROSITA} finds fairly smooth edges of the EBs, not matching the strong deformations seen in this bubble model. This model therefore deviated from the observational EBs and FBs enough to not be considered for further investigations with more realistic TDE rates. However, we note that above a cut-off wavenumber, the strong RT instabilities could be inhibited by magnetic fields \citep{Breitschwerdt2000}, which are not considered in our model.

\subsection{\texorpdfstring{$\beta$-model halo}{Beta-model halo}} \label{sec:results_beta}
  For the $\betaup$-model halo, more realistic TDE rates of $\unit[10^{-4}]{yr^{-1}}$ and $\unit[10^{-5}]{yr^{-1}}$ were considered. The numerical result of the higher TDE rate is shown in the top-right panel of Fig.~\ref{fig:results}. In a comparison between the forward shock of the simulated bubbles and the X-ray EBs of \citet{Predehl2020}, one can see that the spherical and symmetric shape in the northern and southern Milky Way halo is well matched. The shell has fewer and weaker RT instabilities than the exponential halo model bubble because its density gradient is not as strong. It therefore resembles the observations with unperturbed edges better.

  Furthermore, the bubbles grow for \unit[16]{Myr} until they reach the size of the EBs in a $\betaup$-model halo, leading to a total injected energy of \unit[1.5$\times$10$^{56}$]{erg}. A comparison to the actual evolution time of the EBs is difficult since it is highly uncertain. Predictions range from a few Myr for an AGN jet model \citep{Yang2022} to $\sim$\unit[20]{Myr} for either a different AGN jet model with a circumgalactic medium wind \citep{Mou2023}, or theoretical calculations with today's shock velocity \citep{Predehl2020} and the Kompaneets formalism \citep{Schulreich2022}. The evolution time of our model agrees well with the upper limit of this range. We note that TDEs can also happen before the derived evolution time in our model as the lifetime of Sgr~A$^*$ is much longer. However, as stars typically gather in clusters, TDEs often occur concurrently. Thus, a quiescent phase with no Galactic outflow and thus no bubbles, followed by an episode of successive TDEs could explain why the EBs and FBs only started to be driven a few tens of Myr ago, and were not present before. Since the real times of past TDEs are unknown, we chose to model the simplest possibility, meaning that a TDE quiet period ended \unit[16]{Myr} ago and was followed by a phase of continuous TDEs.

  The higher TDE rate $\betaup$-model halo bubbles still had potential for improvement because they exhibited quite strong RT instabilities at the contact discontinuity that almost penetrated the shell although being weaker than in the exponential halo model. This would result in $\gammaup$-ray emission at very high latitudes, different to the FBs which only reach up to $\pm 50^\circ\text{ to} \pm60^\circ$. Thus, the model could still be improved.

  The bottom-left panel of Fig.~\ref{fig:results} shows the results of the lower realistic TDE rate ($\unit[10^{-5}]{yr^{-1}}$) simulation. No strong deformation of the contact discontinuity by RT instabilities is visible anymore, since there are fewer but stronger events. The forward shock shape hardly changes because the total injected energy remains the same.
  
\begin{figure*}[hbpt!]
      \centering
      \begin{minipage}[t]{\linewidth}
      \includegraphics[width=.98\linewidth]{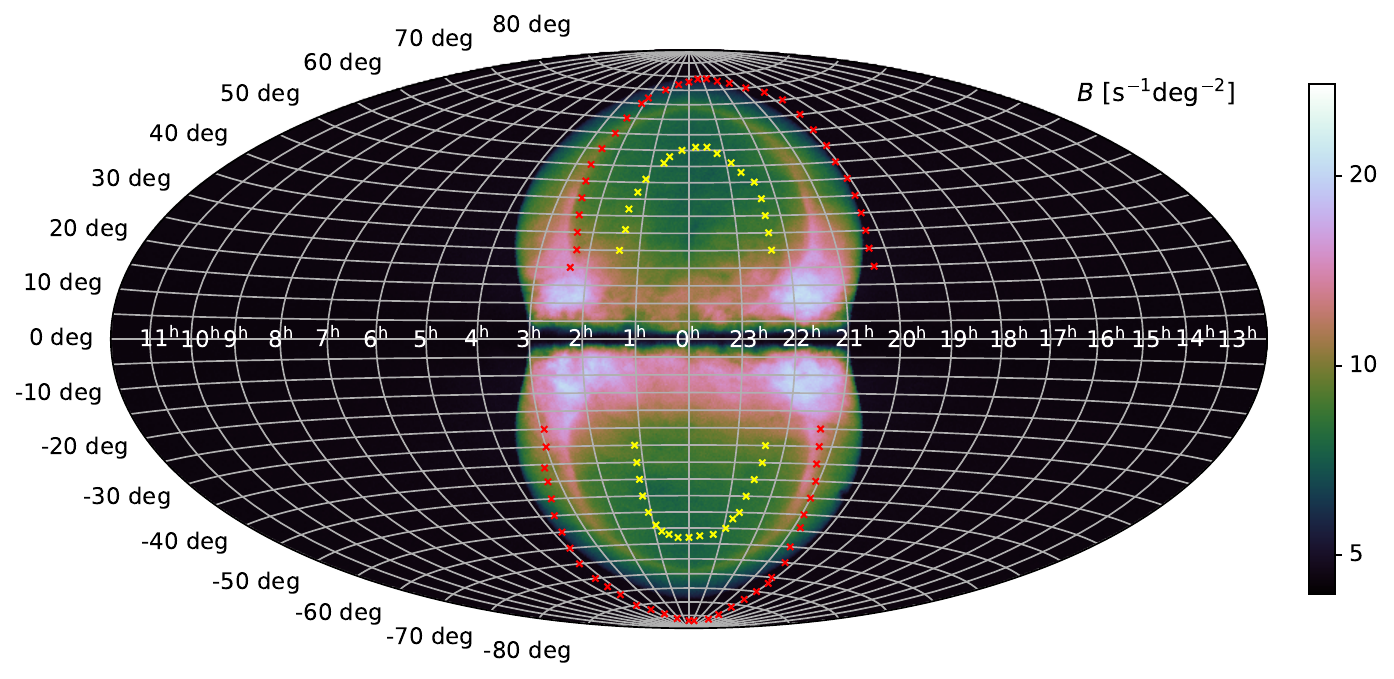}
      \end{minipage}
      \begin{minipage}[t]{\linewidth}
      \includegraphics[width=.98\linewidth]{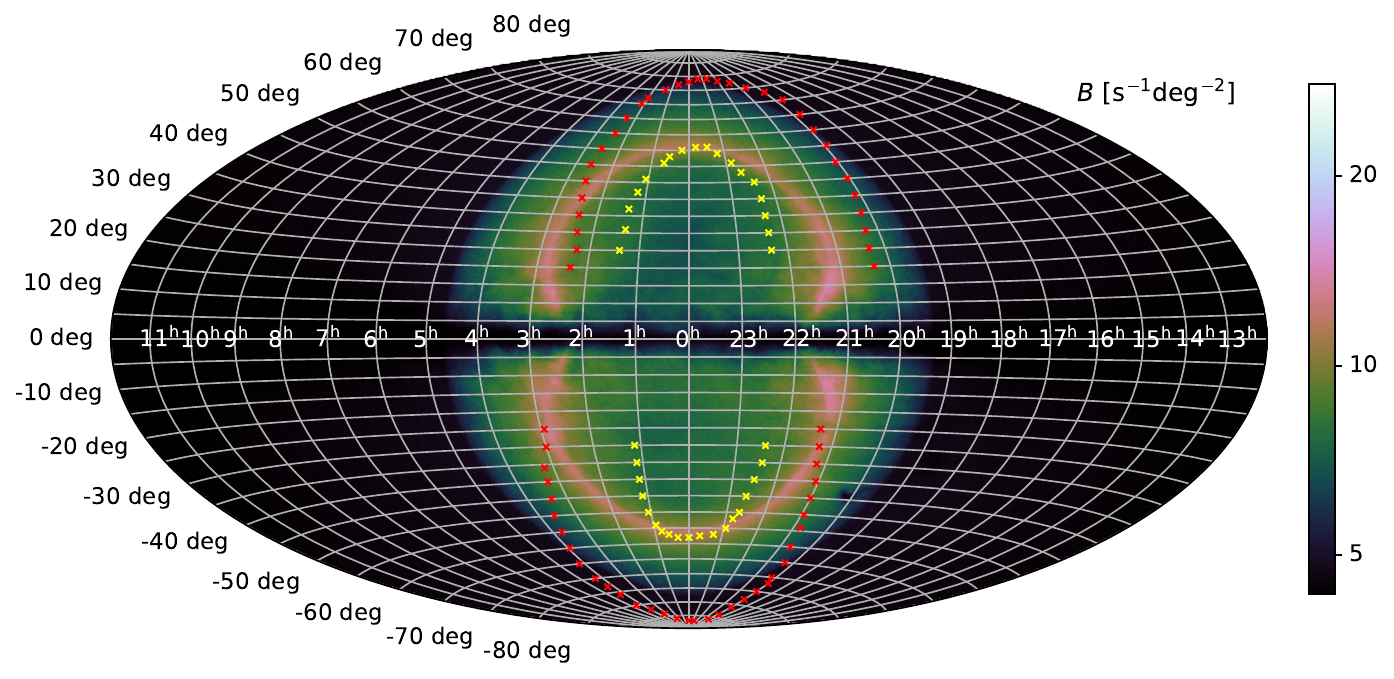}
      \end{minipage}
      \caption{Synthetic observations of the full three-dimensional simulation results using a Hammer-Aitoff projection in Galactic coordinates. The color-coded intensity is the surface brightness $B$ in logarithmic units for the photons detected per time and solid angle and corresponds to detected X-ray photons with an energy between \unit[0.6]{keV} and \unit[1]{keV}. Since the cosmic X-ray background was not included in our model, we added it as a constant value to the maps so that we arrive at the same minimum brightness of \unit[4.66]{photons s$^{-1}$ deg$^{-2}$} as \citet{Predehl2020}. The colorbar limits \citep[obtained from][]{Tseng2024} are the same as in the observations of \citet{Predehl2020}. The main underlying model is the thermal collisional ionization equilibrium model of \texttt{Cloudy}. Both, the geometry for an observation from the Solar System and instrumental effects of the \textit{eROSITA} telescope were considered for these figures. The FBs would appear at the contact discontinuity, which lies at the inner side of the brighter region (pink-white color) according to our model. The approximate shell of the observed EBs (red) and FBs (yellow) are indicated using data points obtained from \citet[][Fig. 3]{Predehl2020}. We note that these points have a reading error of about $\pm5^\circ$ and $\pm\unit[1]{h}$ because the edges of the observed EBs and FBs are not clearly defined everywhere. The top panel shows the $\betaup$-model halo bubble (cf.~Fig.~\ref{fig:results} bottom left) and the bottom panel shows the multicomponent model halo bubble (cf.~Fig.~\ref{fig:results} bottom right).}
      \label{fig:synthetic_obs}
  \end{figure*}

  To better compare the two-dimensional slices to the \textit{eROSITA} and \textit{Fermi} all-sky surveys, we calculated synthetic X-ray maps from the three-dimensional simulation results most similar to the EBs and FBs and plotted them in a Hammer-Aitoff projection using Galactic coordinates, analogous to \citet{Predehl2020}. Because our model did not include background X-ray radiation, we added it as a constant value, so that the minimum brightness arrived at \unit[4.66]{photons\,s$^{-1}$\,deg$^{-2}$}. This value represents the mean of the observed brightness profiles \citep[][Fig.~2b]{Predehl2020} above $\alpha_\text{G}=90^\circ$ and below $\alpha_\text{G}=-90^\circ$. The map of the $\betaup$-model halo bubble with a TDE rate of $\unit[10^{-5}]{yr^{-1}}$ is shown in the top panel of Fig.~\ref{fig:synthetic_obs}. Since our model does not consider CRs, we are not able to obtain synthetic $\gammaup$-ray observations. We do, however, argue that the $\gammaup$-ray emission should occur at the contact discontinuity of the simulated bubbles. This would appear in Fig.~\ref{fig:synthetic_obs} at the inner edge of the brighter (pink-white) region. The EB and FB observations of \citet{Predehl2020} and \citet{Su2010} are indicated with red and yellow data points, respectively. One can see that the overall structure is well matched but the simulated forward shock and especially the contact discontinuity are a bit too wide.
  
  There are several reasons for this. First of all, the actual times of past TDEs in the Milky Way are unknown. There could have occurred multiple TDEs concurrently or subsequently, or no TDEs for a longer period of time, altering the evolution and possibly the shape of the contact discontinuity. Second, the energy and mass of the past TDEs are unknown. If the TDEs had an higher overall luminosity than assumed, the bubbles would need less time to spread into the halo, possibly leading to a narrower forward shock and contact discontinuity, better resembling both superbubbles. Third, the Milky Way temperature and density distribution are highly uncertain. We try to overcome this by considering different Milky Way models, however, all models still may be too simplistic to match the actual Milky Way density distribution, lacking clumps or other inhomogeneities that can alter the forward shock evolution. Fourth, although neglecting cooling and the effects of magnetic fields may be justified for calculating the overall structure of the bubbles, their inclusion will certainly have an influence locally and on the detailed structure. Taking all of these limitations into account, the simulation results match the observations fairly well.

\begin{figure*}[hbpt!]
      \centering
      \begin{minipage}[t]{0.95\linewidth}
      \includegraphics[width=\linewidth]{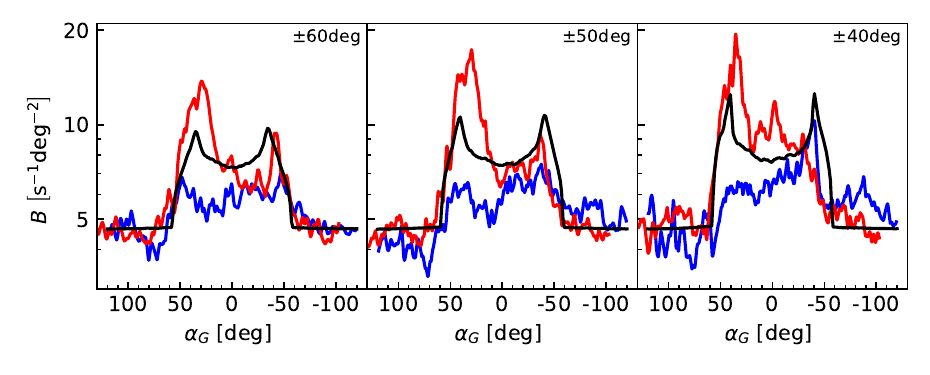}
      \end{minipage}
      \begin{minipage}[t]{0.95\linewidth}
      \includegraphics[width=\linewidth]{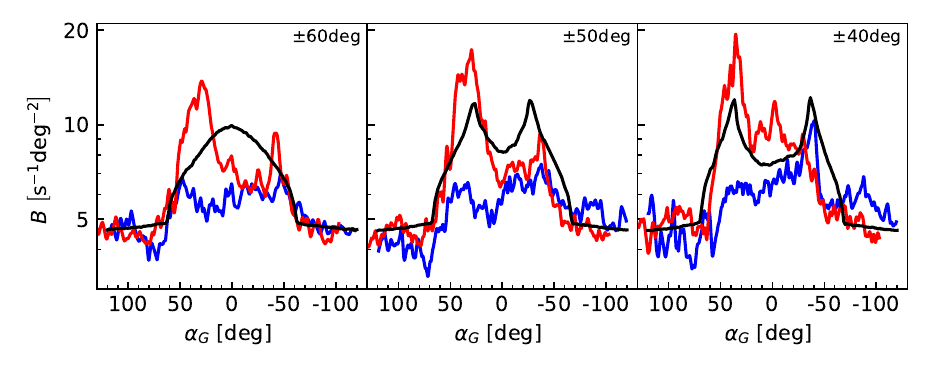}
      \end{minipage}
      \caption{Brightness profiles of our synthetic observations at different latitudes. The top panel shows the surface brightness $B$ for the $\betaup$-model halo bubble and the bottom panel for the multicomponent model bubble for the latitudes $\pm40^\circ$, $\pm50^\circ$, and $\pm60^\circ$ along the longitudes $\alpha_G$. Only the northern brightness profiles of our synthetic observations (black) are shown and compared to the data of the northern (red) and southern (blue) EB from \citet{Predehl2020}. The northern \textit{eROSITA} data was shifted by $16.5^\circ$ to the east. Furthermore, a constant background X-ray radiation was added to our data similarly to Fig.~\ref{fig:synthetic_obs} to better compare it to the observations.}
      \label{fig:brightness}
  \end{figure*}

  This is supported by the brightness profiles of our synthetic X-ray maps, shown in the top panel of Fig.~\ref{fig:brightness}. At latitudes $\pm40^\circ$, $\pm50^\circ$, and $\pm60^\circ$, these can be compared to the \textit{eROSITA} data in \citet[][Fig.~2b]{Predehl2020}, which are shown in Fig.~\ref{fig:brightness} in red and blue for the northern and southern EB, respectively. As our bubbles are almost perfectly symmetric with respect to the Galactic plane, we refrain from showing the southern latitudes separately. The background radiation was again added as a constant value, similarly to Fig.~\ref{fig:synthetic_obs}. The observed northern EB data was furthermore shifted by $16.5^\circ$ to the east to account for the observed EB shift. In a comparison of the brightness profiles, our simulations and the observations agree fairly well. At all shown latitudes, the brightness profiles of our synthetic $\betaup$-model halo bubbles and especially the northern EB have peaks at $\pm40^\circ$ longitude. Since our model does not consider any asymmetries, the eastern peak of the observations is brighter and wider than in our maps because it is influenced by the very bright North Polar Spur, which lies on the outer edge of the EBs. This is also why the bright region in the east appears to be more on the outside in the observations. The western peak, without North Polar Spur, is slightly dimmer but similarly wide compared to our model. A decrease in brightness between those peaks can be seen in the northern EB and in our data. As the observed southern EB is not very bright, a detailed comparison to it is more difficult. The brightness increase of the southern EB, however, still seems to be at similar longitudes as the northern EB and our synthetic observations. Compared to the brightness profiles of synthetic X-ray maps of other studies that use an AGN jet model, such as \citet{Yang2022}, \cite{Mou2023}, \citet{Chang2024} and \citet{Tseng2024}, our brightness profiles seem to match fairly well the \textit{eROSITA} data, especially in terms of peak visibility and position. Of course we note that all numerical models still had many initial assumptions that have a significant impact on the brightness.

  Since there is no termination shock visible in the resulting bubbles (bottom-left panel of Fig.~\ref{fig:results}) due to the low TDE rate, we furthermore wanted to study whether the TDEs lead to a shock at all. Otherwise, the $\gammaup$-ray emission of the FBs may be left unexplained as CRs could not be sufficiently accelerated. To study this in detail, we applied a shock tracer algorithm. Following a similar approach as \citet{Schaal2015} and \citet{Pfrommer2017}, we found shocked cells by checking if
  \begin{enumerate}[(i)]
      \item $\Vec\nabla \cdot \Vec \varv < 0$,
      \item $\Vec \nabla T \cdot \Vec \nabla \rho >0$, \text{and}
      \item $\mathcal{M} > 1.0$.
  \end{enumerate}

  These conditions were checked for a central simulation box slice, also considering the adjacent slices to allow for a correct evaluation of the finite differences in all spatial directions. Figure~\ref{fig:shock_betar} (provided in the appendix) shows the detected shocked cells for the TDE shock evolution marked in gray. The snapshots were taken from \unit[18--18.1]{Myr}, slightly after the observed size of the EBs had been reached.

  \begin{figure}[hbpt!]
      \centering
      \includegraphics[width=.96\linewidth]{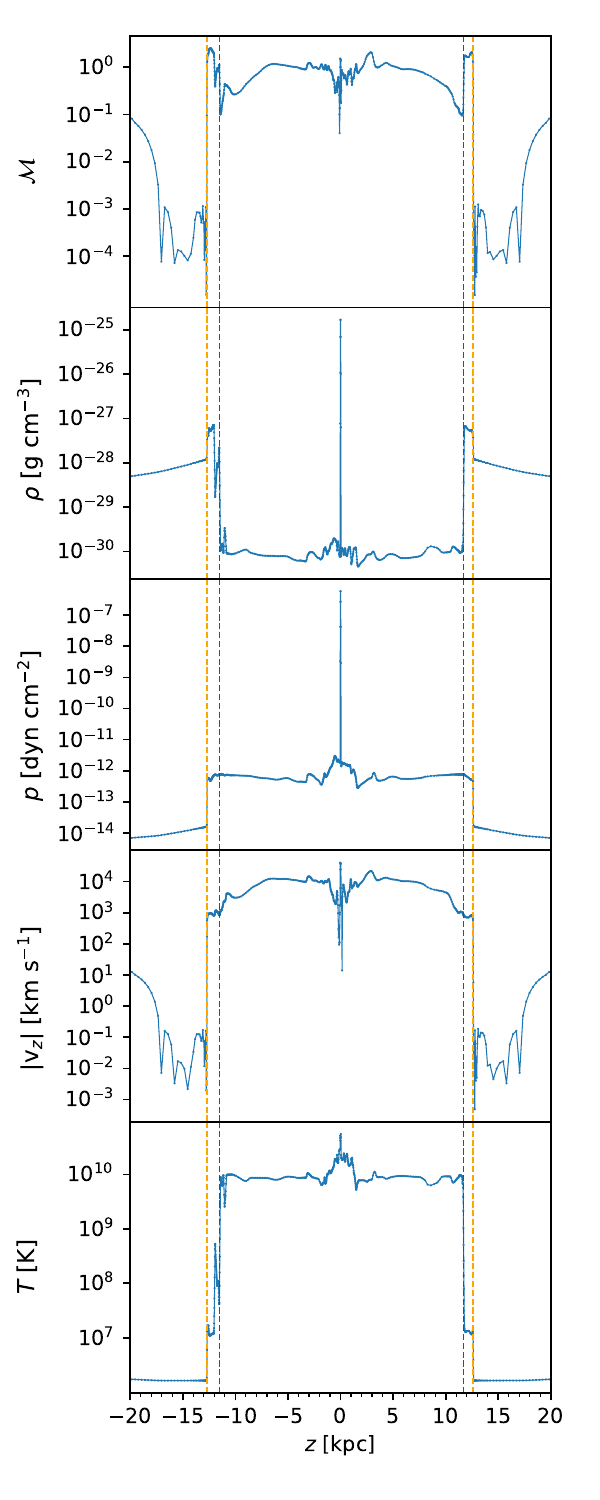}
      \caption{Simulated vertical profiles of  the Mach number $\mathcal{M}$, the density $\rho$, the pressure $p$, the absolute value of the flow speed in $z$-direction $v_z$, and the temperature $T$ measured along a ray through the origin (injection region) of TDEs occurring at a rate of $\unit[10^{-5}]{yr^{-1}}$ in the $\betaup$-model halo at \unit[16]{Myr}. The green and orange dotted lines indicate the position of the contact discontinuity and the forward shock, respectively.}
      \label{fig:1d_betar_100kyr}
  \end{figure}
  
  One can see in Fig.~\ref{fig:shock_betar}(a) that the newly injected TDE creates a shock front in the very center of the slice. In addition, ellipsoidal shock fronts from previous TDEs are still visible, extending to at least \unit[4]{kpc} in the slice shown. The new TDE shock then propagates rapidly through the interior due to its high energy and therefore catches up and collides with a slower pre-existing shock in (b). As a result of the collision, the TDE shock splits into a transmitted shock, which continues to propagate toward the shell, and a reverse shock, which moves back to the center. In (c)--(e), the forward shock continues on its path only slower, whereas the reverse shock reaches the center in (e) and reverses a second time. In (f), both shocks still move outwards, while a new TDE shock is already traced in the center of the box because \unit[100]{kyr} have passed since the last TDE.

  Although we showed that there are shocks traveling through the interior of the bubble, their Mach number of $\mathcal{M}=2$--$3$ (Fig.~\ref{fig:1d_betar_100kyr}, top panel) might be too low to accelerate CRs to high enough energies. However, our model assumed one of the worst cases for building high-Mach TDE shocks, since the TDEs all had the same energy, and happened only every \unit[100]{kyr}. Therefore, stronger shocks due to occasional higher energy TDEs or consecutive TDEs from stellar clusters were never encountered in our model. Furthermore, once the bubbles have expanded, it is to be expected that shock waves due to ordinary SN explosions will easily travel out into the halo, catching up with previous shocks, reinforcing them, and strengthening as they travel down a density gradient, as has been proposed in the context of CR-driven galactic winds \citep{DB:12}.

  Figure~\ref{fig:1d_betar_100kyr} also shows the velocity of the shocks. The TDE shocks reach a vertical speed of \unit[$\sim$10$^4$]{km\,s$^{-1}$}, while the velocity is an order of magnitude lower for the forward shock. The recently injected TDE has effects on all fluid quantities shown and can therefore be easily identified in all panels at $z=$ \unit[0]{kpc}. Other properties of our results show that the interior has a low density, a high temperature, and is almost isobaric, whereas the shell has a high density, a high Mach number, and a lower temperature than the interior, perfectly aligning with typical properties of superbubbles.
  
  For a further comparison to the EBs, the shell thickness of the simulated superbubble, which is the separation distance between the forward shock (dashed orange line) and the contact discontinuity (dashed green line), can be determined from the density, Mach number, and temperature profile to be $\gtrsim$\unit[1]{kpc}. \citet{Predehl2020} estimate the EB shell to be $\sim$\unit[2]{kpc} thick by comparing geometric models to surface brightness maps at different Galactic latitudes. The synthetic observations of Fig.~\ref{fig:synthetic_obs} also support the conclusion that the shell is too thin in our model and thus, the FBs are too large. However, as stated earlier, we have not considered magnetic fields in our simulations, which would certainly increase the shell thickness because magnetic pressure would reduce the overall compression.
  The real magnetic field in the Milky Way and especially the Galactic center is not known with certainty as it is rather complex \citep[see review of][]{Beck2013}. However, poloidal components \citep[e.g.][]{Ferriere2009} could be swept up by the TDE shocks in our model, leading to a thicker shell in both directions, parallel and perpendicular to the Galactic plane, which would resemble the FBs better. Other changes such as a slightly different density distribution or a non-constant TDE rate could also impact the shell thickness significantly. We leave the detailed investigation of the impact of these different parameters for a future TDE study.

  To summarize, we find that our $\betaup$-model halo bubbles reproduce the main features of the EBs and FBs fairly well, given the numerous unknown and uncertain initial parameters. Furthermore, we traced TDE shocks in the interior that could explain the CR $\gammaup$-ray emission. However, the FB shape does not perfectly match this model. An improvement can be made by including magnetic fields that would increase the shell thickness, or occasional high-energy or clustered TDEs that would affect the inner structure significantly. Furthermore, more realistic simulations could also include short-lived collimated jets that are observed for some TDEs \citep[e.g. Swift 1644+57/GRB 110328, see][]{Bloom2011, Shao2011}. These jets would induce a direction to the rather omnidirect unbound stellar debris of the TDEs and thus lead to a more ellipsoidal FB shape.


\subsection{Multicomponent Milky Way model} \label{sec:results_mcmillan}
  Since the lower TDE rate of $\unit[10^{-5}]{yr^{-1}}$ did not change the bubble structure significantly in the $\betaup$-model but resembled the FBs slightly better than the higher TDE rate model due to fewer RT instabilities, the multicomponent model was studied with the lower TDE rate only. The simulation results are shown in the bottom-right panel of Fig.~\ref{fig:results}. The shape of the bubbles differs significantly from the previously presented models as the Milky Way density gradient does not decrease as steeply in the $z$-direction in this model. The resulting bubbles are smaller but wider when they reach the size of the EBs. They also show no constriction in the mid-plane, thus not matching well with the EBs that are more spherical. The TDEs take \unit[14.5]{Myr} to form the superbubbles in this model, totaling an energy injection of \unit[1.4$\times$10$^{56}$]{erg}, similar to the $\betaup$-model. 
  
  Once again we produced a synthetic X-ray map, which is shown in the bottom panel of Fig.~\ref{fig:synthetic_obs}. It emphasizes that the simulated superbubbles in a multicomponent Milky Way model are too wide at the mid-plane, spanning over \unit[4]{h} in longitude, while being as large as the EBs (red) in latitude. This can be seen once again in the brightness profiles in Fig.~\ref{fig:brightness}, especially at $\pm50^\circ$ and $\pm60^\circ$ latitude, where the brightness peaks are not that far separated or not separated at all, except at low latitudes as seen at $\pm40^\circ$, where it is similar to the $\betaup$-model. If the multicomponent model superbubbles grew larger, brightness peaks would also appear in the higher latitude data, potentially aligning better with the observations. Despite that, the peaks at $\pm40^\circ$ would separate further and thus might no longer match observations. Therefore, the previous $\betaup$-model halo bubbles seem to reproduce the EBs better. Furthermore, the contact discontinuity in this model, which lies at the inner edge of the brighter region in Fig.~\ref{fig:synthetic_obs}, is too wide, but shows a similar size in latitude to the FBs (yellow). However, it does not reproduce the FBs much better than the $\betaup$-model.
  
  \begin{figure}[hbtp!]
      \centering
      \includegraphics[width=.96\linewidth]{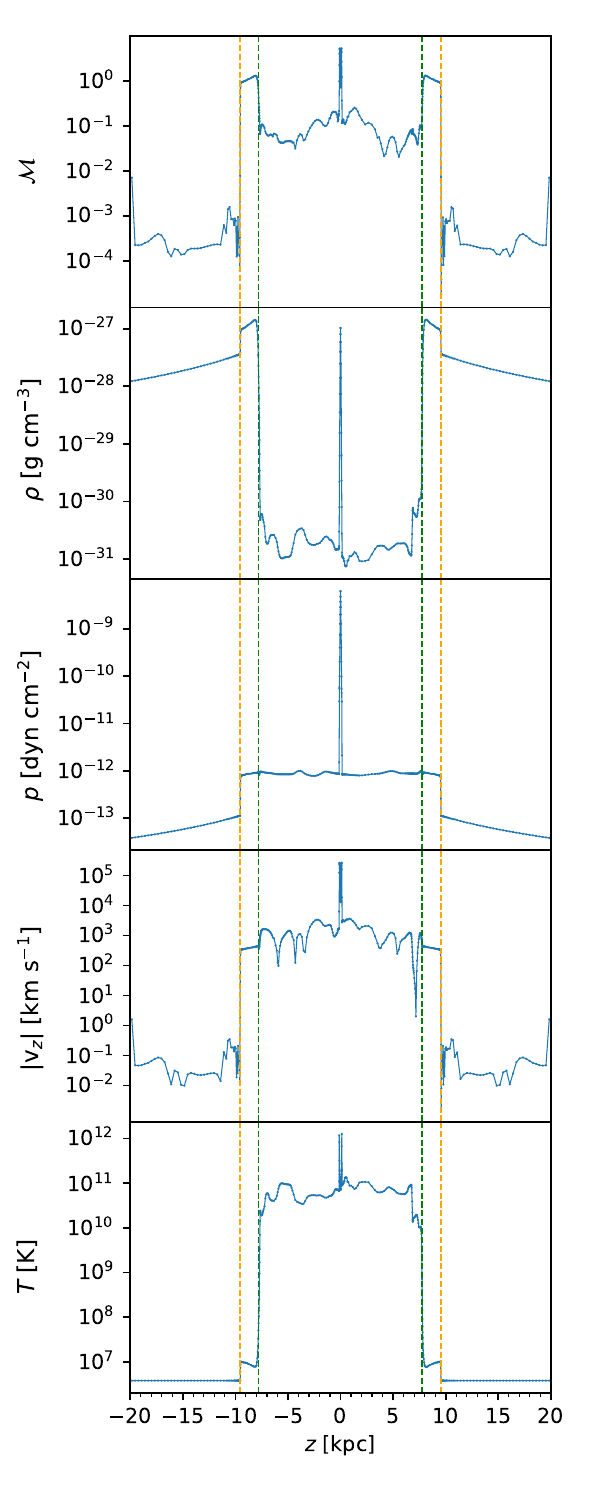}
      \caption{Similar to Fig.~\ref{fig:1d_betar_100kyr} but for the multicomponent model.}
      \label{fig:1d_mc_100kyr}
  \end{figure}

  The shock propagation is very similar to the TDE shocks in the $\betaup$-model halo, which is why we refrain from presenting the shock evolution once again. The strength of the shocks, however, is different from the previous models since the density in the interior of the bubbles is different. That is why we still show one-dimensional plots through the $x=y=0$ line in Fig.~\ref{fig:1d_mc_100kyr} for this model, including the Mach number. Compared to the previous $\betaup$-model plots, the quantities do not have strong gradients in the shell, as no RT instability formed in the contact discontinuity (Fig.~\ref{fig:results}). The bubbles also have a higher temperature and a lower density. The TDE shocks start with a very high Mach number of almost $\mathcal{M}=10$ but also lose their energy more quickly -- shocks are only found in the very center of the superbubbles. Therefore, the $\gammaup$-rays would need to be accelerated more centrally by the initial high Mach TDE shock. The vertical speed of the forward shock is with $\sim$$\unit[500]{km\,s^{-1}}$ also lower in this model. The interior of the bubbles again has a low density, a high temperature, and is almost isobaric, whereas the shell has a high density and a low temperature. The shell thickness was estimated from the density, the Mach number, and the flow speed to be \unit[2]{kpc}, better matching the estimates of \citet{Predehl2020}.

  Concluding, the bubbles blown in the multicomponent model with our setup do not match any better with the EBs and FBs than the bubbles in the $\betaup$-model do. Even though the shell is thicker and the FBs therefore smaller, the simulated FBs and EBs are too wide and the forward shock is not spherical.

  On a different note, the asymmetric shape of the EBs remains without any explanation in our investigations since our set-up is axisymmetric and has a symmetry with respect to the Galactic plane. An asymmetric density distribution or a postulated circumgalactic wind \citep{Mou2023} may be able to explain this property. Although \citet{Mou2023} use an AGN outflow as energy source, their considered outflow has similarities to the thermal energy outflow produced by the TDEs in our model since it has a constant \unit[19]{Myr}-long low luminosity. The evolution of a TDE blown bubble with asymmetric features needs to be investigated in detail in future studies.

\section{Conclusions} \label{sec:conclusions}
  Three-dimensional hydrodynamic simulations using a TDE model were carried out to reproduce the EBs and the FBs. Our findings can be summarized as follows.
  \begin{enumerate}
      \item Regular energy injections with a rate of \unit[$10^{-5}$]{yr$^{-1}$} and an average luminosity of \unit[3$\times$10$^{41}$]{erg\,s$^{-1}$} at the Galactic center by for example TDEs can blow superbubbles into the Galactic halo, having a comparable spherical and smooth shape, size, and evolution time to the EBs. Simulations with a $\betaup$-model halo reproduce the typical EB properties better than simulations with an exponential halo or multicomponent Milky Way model.
      \item The general cocoon-like structure of the EBs around the FBs can be reproduced by our model. We predict that the $\gammaup$-ray emission occurs at the contact discontinuity of the simulated bubbles.
      \item Tidal disruption events create shock fronts inside the bubbles that may be able to continuously accelerate CRs to energies high enough to explain the FB $\gammaup$-rays, be they leptonic or hadronic in origin. The shocks found in our study are not certain to be strong enough to do so but could be strengthened by more consecutive or stronger TDEs, and regular SN explosions in the disk, however, this needs to be investigated further.
      \item Synthetic observations of our simulated superbubbles show that the X-ray emission occurs mostly in the dense shell. Together with the FBs at their contact discontinuity, they confirm the points 1 and 2 once again. 
      \item The TDE model can reproduce the brightness profiles observed with \textit{eROSITA} fairly well. The brightness peaks at about $\pm40^\circ$ longitude with a decrease in brightness between them are visible in both the observations and the synthetic maps. Furthermore, the absolute brightness of our generated observations is quite close to that of the actual observations.
      \item Finally, we would like to stress that the agreement between the observations of the EBs and FBs, and our hydrodynamic models is fairly good, considering the tremendous impact of the detailed local gas density and pressure distributions in the Galactic halo on the shape of the bubbles, as is also well known from the modeling of large superbubbles in the ISM.
  \end{enumerate}

  Nevertheless, some discrepancies between model and observations need further discussion. We note that especially the FBs are not matched perfectly by our model.
  We argue that this offset can be explained by the uncertainty of several input parameters that have a strong impact on the forward shock and contact discontinuity shape. Besides the Milky Way temperature and density distribution, for which we investigate several models but which are still quite unconstrained, the actual times, masses, and energies of past TDEs are unknown. Furthermore, CRs, radiative cooling, self-gravity, and magnetic fields may have an impact on the detailed and local structures including the RT instabilities, whereas the global structure should remain unchanged. Due to these uncertainties, more comprehensive simulations may still deviate from the detailed EB and FB shape. Only a thorough study of a large parameter space, including more Milky Way models, a non-constant TDE rate, a variable TDE energy, and short-lived TDE jets that would induce a direction to the outflow may give better insights particularly into the evolution of the FBs with a TDE model. 
  
  Concluding all findings, we showed that TDEs, which are expected to happen regularly in the Milky Way, or other regular energy injection events at the Galactic center, are a possible origin of the EBs and FBs. The fact that we demonstrated that it is possible to match the observations by our TDE model does not rule out that other mechanisms discussed in the literature contribute to the expansion and the emission. It is quite likely that jets were blown out during episodes of strong accretion onto the SMBH, although not in a continuous fashion. In addition, the star formation rate in the Galactic center is known to be higher than in the solar neighborhood, so that SNe, as already been mentioned, would also add their share.

  \begin{acknowledgements} 
  The authors thank Chung-Ming Ko and Romain Teyssier for providing further information on their past paper and the efficiency of the \texttt{RAMSES} code, respectively. We further want to thank Bert Vander Meulen for the helpful discussions regarding our synthetic X-ray maps, Mattia Pacicco for test simulations of the impact of magnetic fields and useful discussions about the X-ray maps, as well as Nina Sartorio for their valuable suggestions and assistance in refining the manuscript.
  TS acknowledges funding from the European Research Council (ERC) under the European Union’s Horizon 2020 research and innovation programme \mbox{DustOrigin} (ERC-2019-StG-851622).
  Some of the results in this paper have been derived using the \texttt{healpy} and \texttt{HEALPix} package \citep{Gorski2005,Zonca2019}. This work also made use of the \texttt{Python} packages \texttt{NumPy} \citep{Harris2020}, \texttt{Matplotlib} \citep{Hunter2007}, MPI for \texttt{Python} \citep{Dalcin2011}, and the \mbox{analysis} and visualization software \texttt{yt} \citep{Turk2011}.
  \end{acknowledgements}

\begin{appendix}
  \onecolumn
  \section{Shocks inside the superbubble}
  To check whether the injected TDEs lead to shocks that may be able to accelerate CRs, we applied a shock tracer algorithm (see Sect.~\ref{sec:results_beta} for details) to the $\betaup$-model halo bubble, as shown in Fig.~\ref{fig:shock_betar}, and studied how these TDE shocks evolve over time.
  \begin{figure*}[hbpt!]
      \centering
      \begin{minipage}[t]{\linewidth}
      \centering
      \begin{tikzpicture}
      \node at (3.27,3.24){\includegraphics[width=0.33\textwidth]{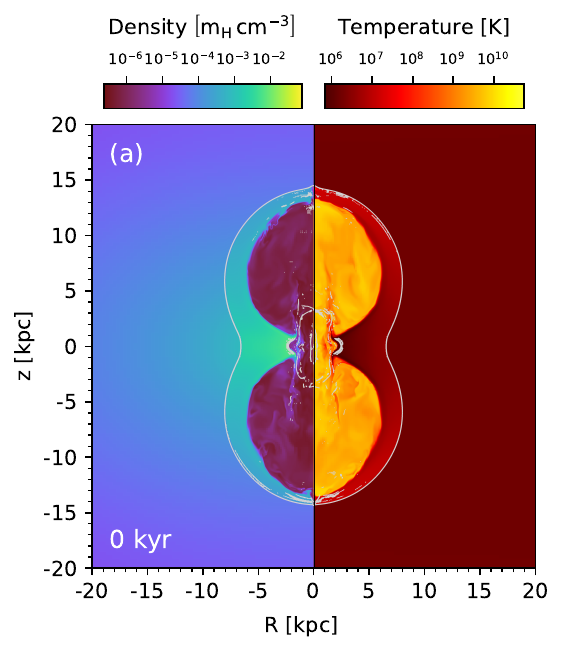}};
      \node at (10,3.24){\includegraphics[width=0.33\textwidth]{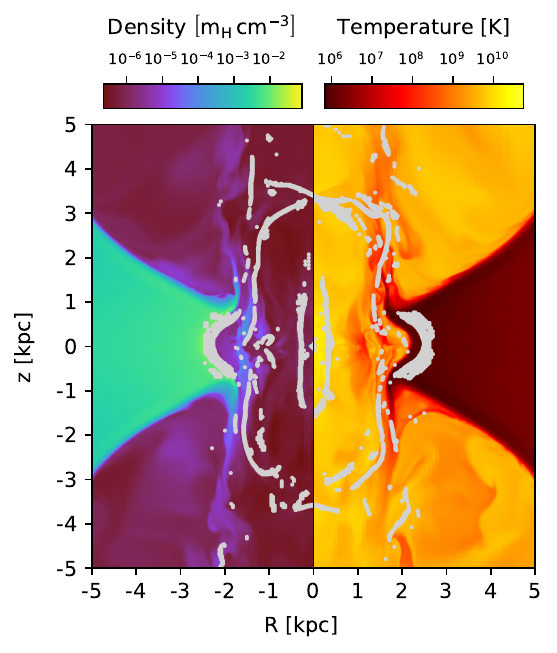}};
      \draw [ultra thick] (3.02,2.43) rectangle (4.22,3.62);
      \draw [ultra thick] (4.22,3.62) -- (7.965,5.425);
      \draw [ultra thick] (4.22,2.43) -- (7.965,0.6);
      \end{tikzpicture}
      \end{minipage}
      \begin{minipage}[t]{\linewidth}
      \vspace{-0.6cm}
      \centering
      \begin{tikzpicture}
      \node at (3.27,3.24){\includegraphics[width=0.33\textwidth]{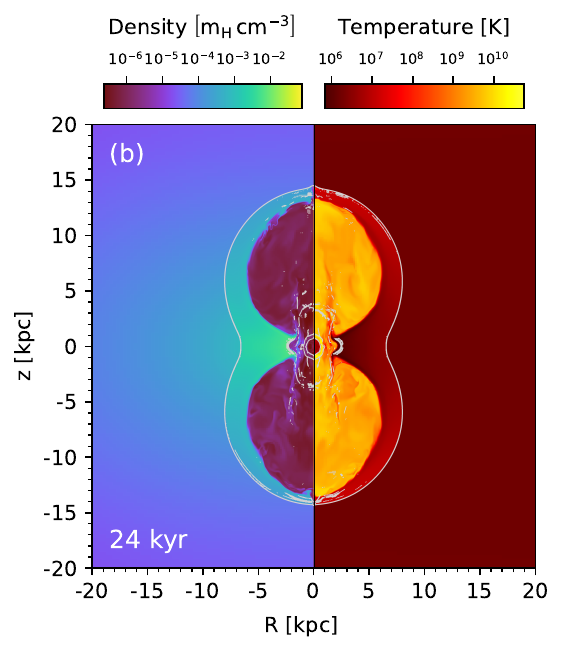}};
      \node at (10,3.24){\includegraphics[width=0.33\textwidth]{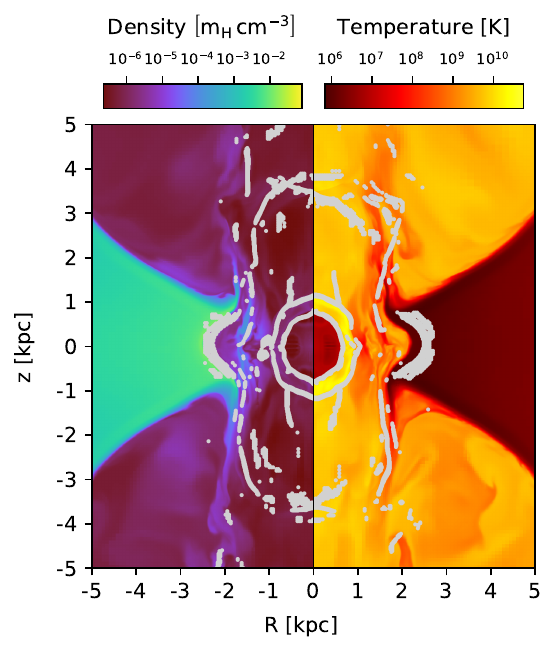}};
      \draw [ultra thick] (3.02,2.43) rectangle (4.22,3.62);
      \draw [ultra thick] (4.22,3.62) -- (7.965,5.425);
      \draw [ultra thick] (4.22,2.43) -- (7.965,0.6);
      \end{tikzpicture}
      \end{minipage}
      \begin{minipage}[t]{\linewidth}
      \vspace{-0.3cm}
      \centering
      \begin{tikzpicture}
      \node at (3.27,3.24){\includegraphics[width=0.33\textwidth]{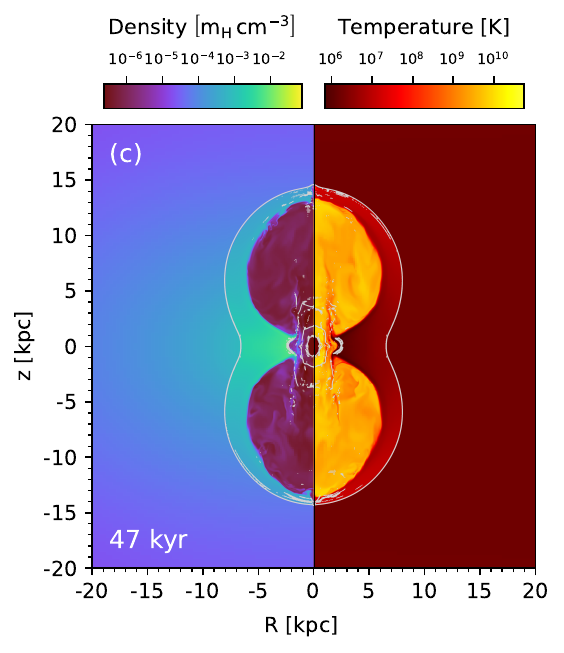}};
      \node at (10,3.24){\includegraphics[width=0.33\textwidth]{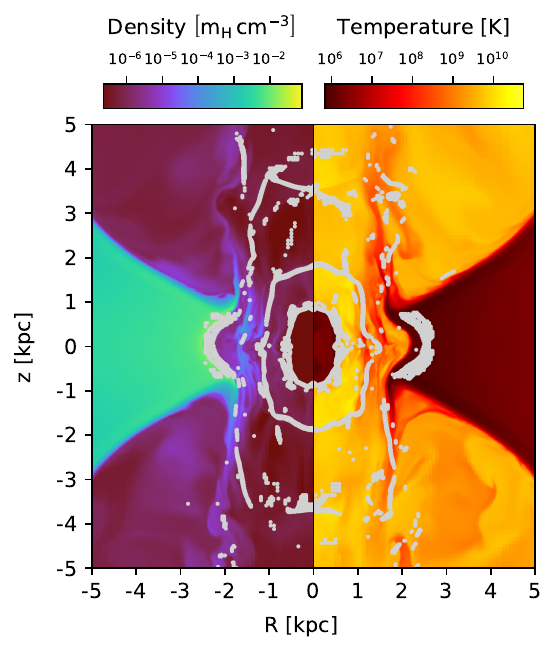}};
      \draw [ultra thick] (3.02,2.43) rectangle (4.22,3.62);
      \draw [ultra thick] (4.22,3.62) -- (7.965,5.425);
      \draw [ultra thick] (4.22,2.43) -- (7.965,0.6);
      \end{tikzpicture}
      \end{minipage}
      \caption{Continued on the next page.}

\end{figure*}
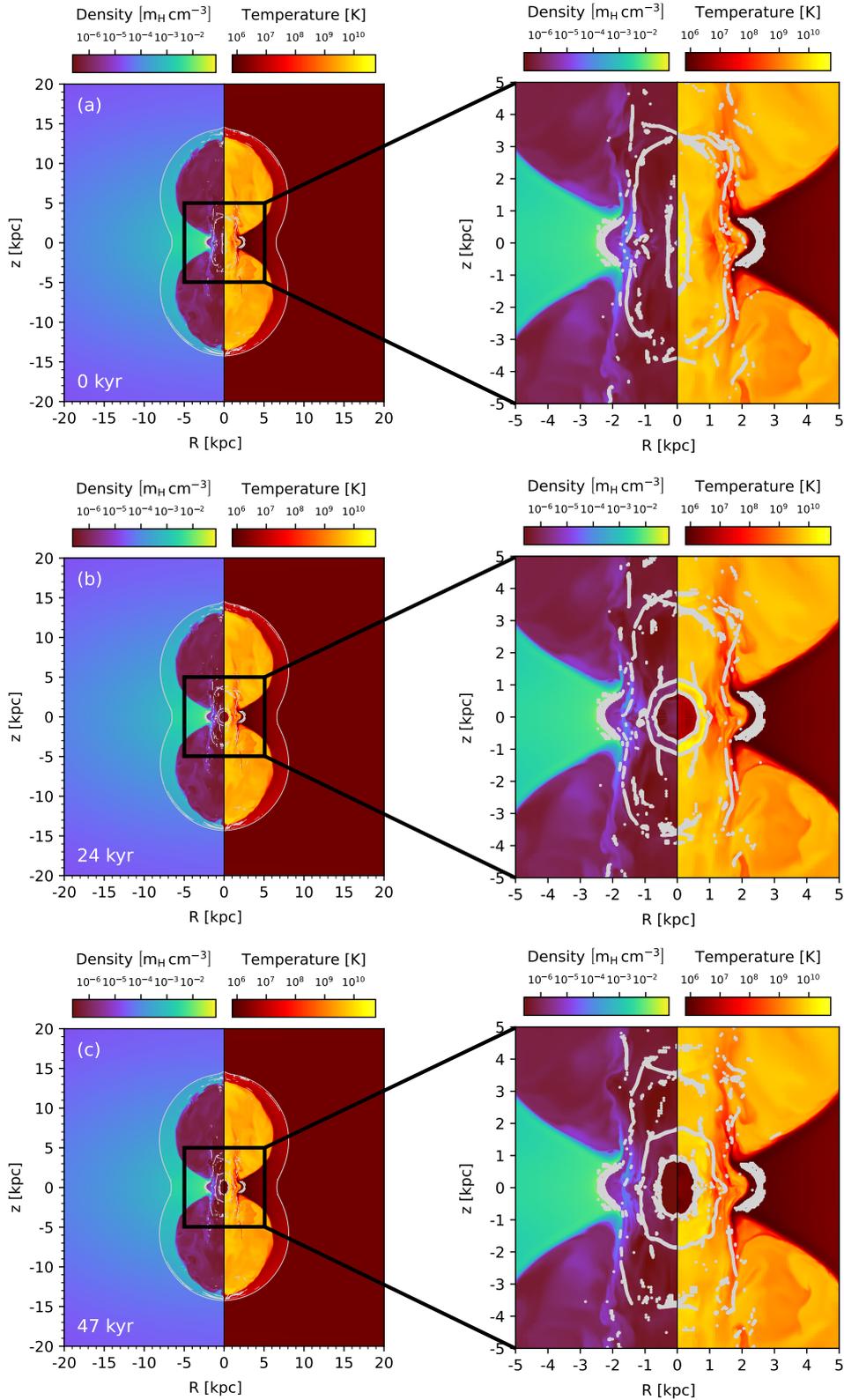
\addtocounter{figure}{-1}
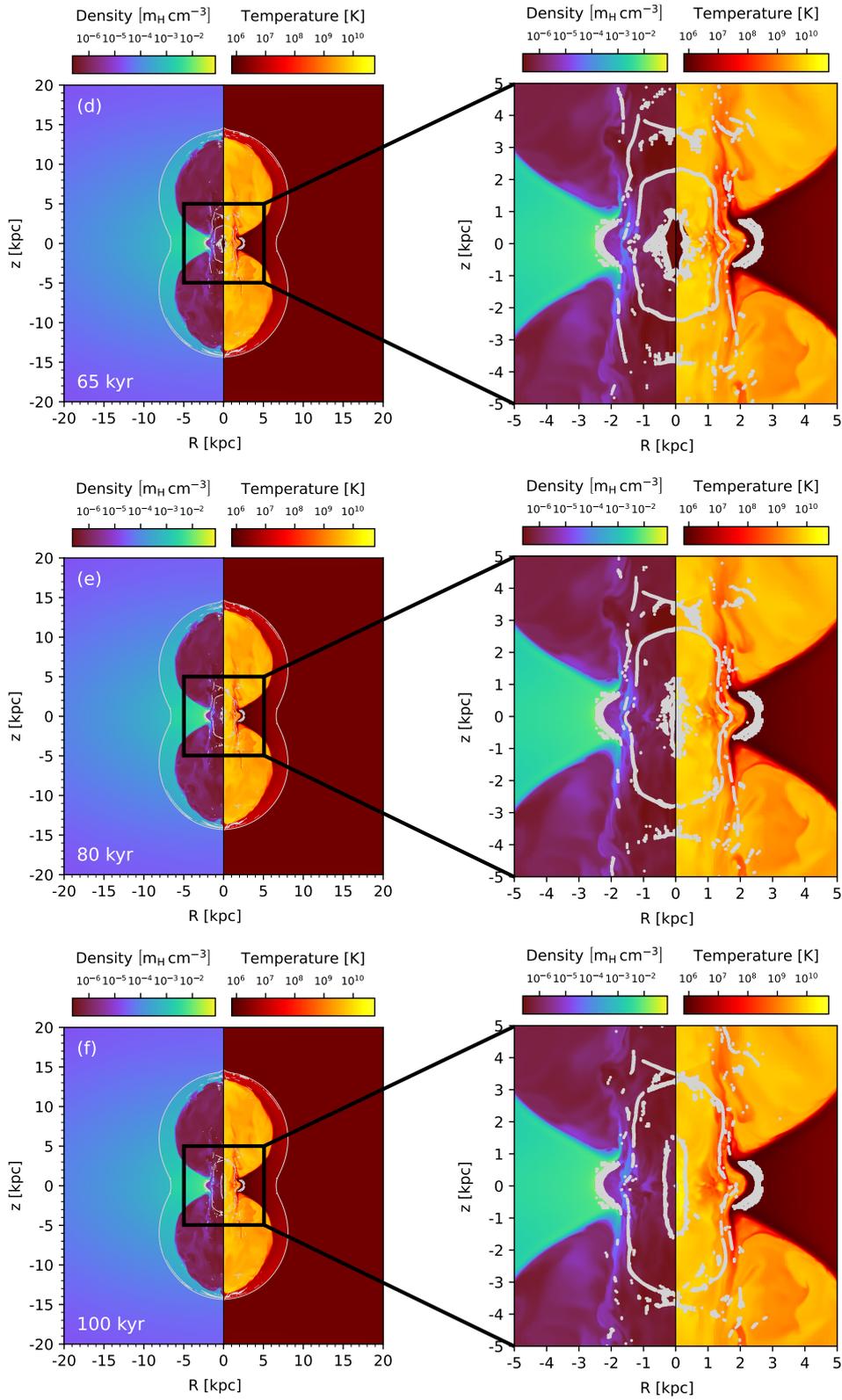
\begin{figure*}
      \begin{minipage}[t]{\linewidth}
      \centering
      \begin{tikzpicture}
      \node at (3.27,3.24){\includegraphics[width=0.33\textwidth]{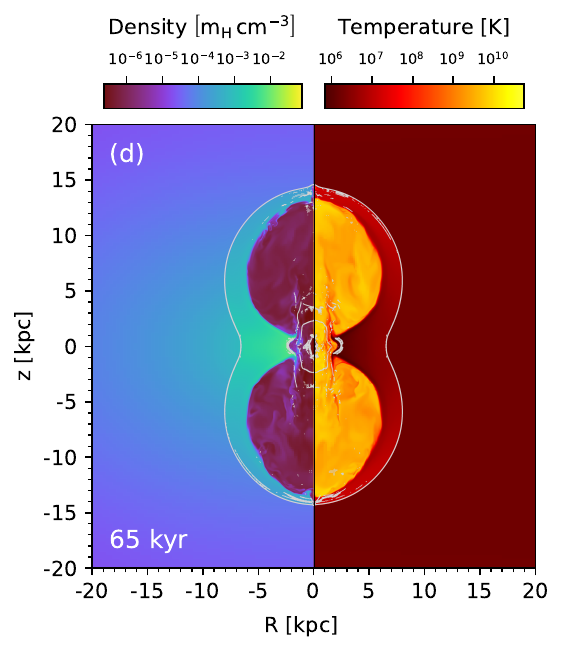}};
      \node at (10,3.24){\includegraphics[width=0.33\textwidth]{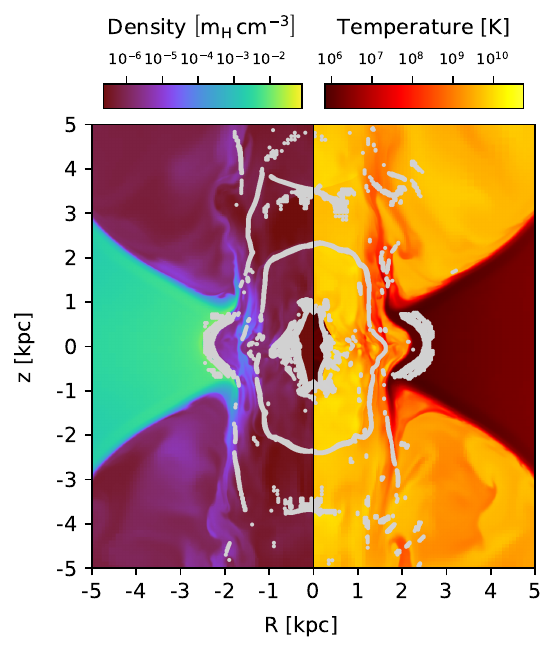}};
      \draw [ultra thick] (3.02,2.43) rectangle (4.22,3.62);
      \draw [ultra thick] (4.22,3.62) -- (7.965,5.425);
      \draw [ultra thick] (4.22,2.43) -- (7.965,0.6);
      \end{tikzpicture}
      \end{minipage}
      \begin{minipage}[t]{\linewidth}
      \vspace{-0.6cm}
      \centering
      \begin{tikzpicture}
      \node at (3.27,3.24){\includegraphics[width=0.33\textwidth]{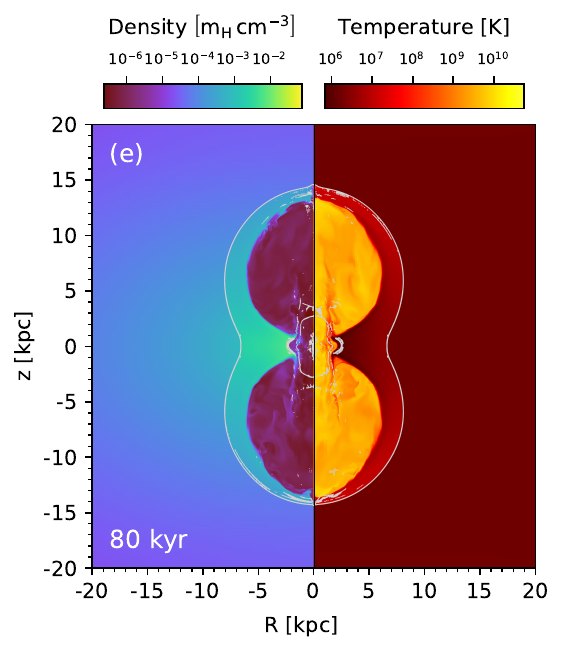}};
      \node at (10,3.24){\includegraphics[width=0.33\textwidth]{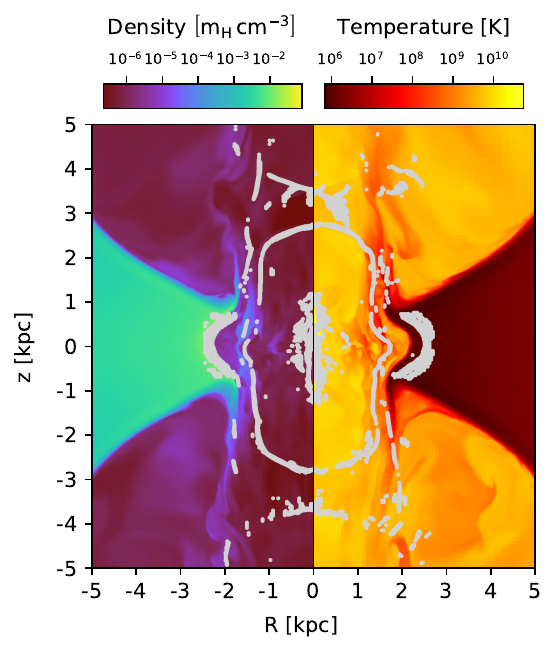}};
      \draw [ultra thick] (3.02,2.43) rectangle (4.22,3.62);
      \draw [ultra thick] (4.22,3.62) -- (7.965,5.425);
      \draw [ultra thick] (4.22,2.43) -- (7.965,0.6);
      \end{tikzpicture}
      \end{minipage}
      \begin{minipage}[t]{\linewidth}
      \vspace{-0.3cm}
      \centering
      \begin{tikzpicture}
      \node at (3.27,3.24){\includegraphics[width=0.33\textwidth]{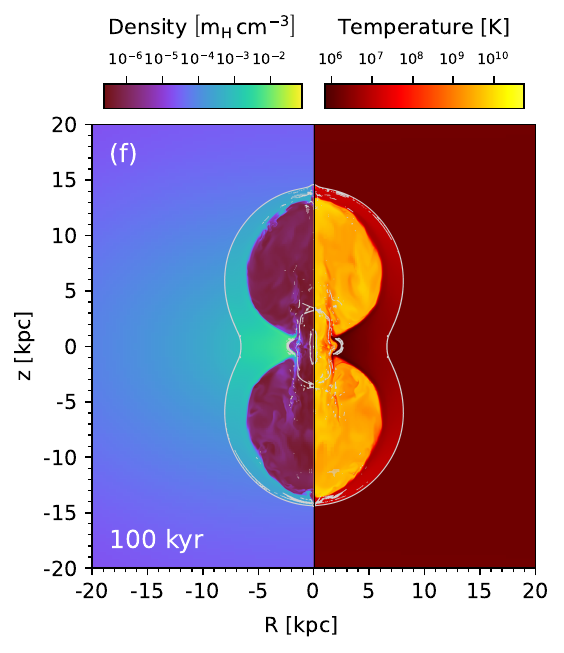}};
      \node at (10,3.24){\includegraphics[width=0.33\textwidth]{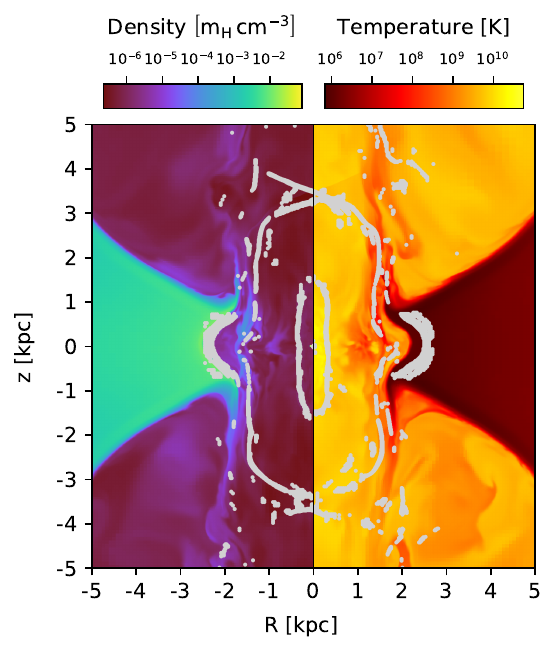}};
      \draw [ultra thick] (3.02,2.43) rectangle (4.22,3.62);
      \draw [ultra thick] (4.22,3.62) -- (7.965,5.425);
      \draw [ultra thick] (4.22,2.43) -- (7.965,0.6);
      \end{tikzpicture}
      \end{minipage}
      \caption{Shock evolution between two TDEs in the $\betaup$-model halo bubble. The shocked cells were found using a shock tracer algorithm and were marked by a gray dot. The right column shows a zoom-in of the inner \unit[5]{kpc} of the left column figures. The elapsed time since the last TDE at \unit[18]{Myr} is indicated. The next TDE happens shortly before snapshot (f).}
      \label{fig:shock_betar}
  \end{figure*} 
\end{appendix}


\begin{thebibliography}{70}
\expandafter\ifx\csname natexlab\endcsname\relax\def\natexlab#1{#1}\fi

\bibitem[{{Abeysekara} {et~al.}(2017){Abeysekara}, {Albert}, {Alfaro}, {et~al.}}]{Abeysekara2017}
{Abeysekara}, A.~U., {Albert}, A., {Alfaro}, R., {et~al.} 2017, \apj, 842, 85

\bibitem[{{Ackermann} {et~al.}(2014){Ackermann}, {Albert}, {Atwood}, {et~al.}}]{Ackermann2014}
{Ackermann}, M., {Albert}, A., {Atwood}, W.~B., {et~al.} 2014, \apj, 793, 64

\bibitem[{{Alexander}(2005)}]{Alexander2005}
{Alexander}, T. 2005, \physrep, 419, 65

\bibitem[{{Axford} {et~al.}(1977){Axford}, {Leer}, \& {Skadron}}]{Axford1977}
{Axford}, W.~I., {Leer}, E., \& {Skadron}, G. 1977, in International Cosmic Ray Conference, Vol.~11, International Cosmic Ray Conference, 132

\bibitem[{{Baumgartner} \& {Breitschwerdt}(2013)}]{Baumgartner2013}
{Baumgartner}, V. \& {Breitschwerdt}, D. 2013, \aap, 557, A140

\bibitem[{{Beck} \& {Wielebinski}(2013)}]{Beck2013}
{Beck}, R. \& {Wielebinski}, R. 2013, in Planets, Stars and Stellar Systems. Volume 5: Galactic Structure and Stellar Populations, ed. T.~D. {Oswalt} \& G.~{Gilmore}, Vol.~5 ({Springer, Dordrecht}), 641

\bibitem[{{Bell}(1978)}]{Bell1978}
{Bell}, A.~R. 1978, \mnras, 182, 147

\bibitem[{{Blandford} \& {Ostriker}(1978)}]{Blandford1978}
{Blandford}, R.~D. \& {Ostriker}, J.~P. 1978, \apjl, 221, L29

\bibitem[{{Bloom} {et~al.}(2011){Bloom}, {Butler}, {Cenko}, \& {Perley}}]{Bloom2011}
{Bloom}, J.~S., {Butler}, N.~R., {Cenko}, S.~B., \& {Perley}, D.~A. 2011, GRB Coordinates Network, 11847, 1

\bibitem[{{Breitschwerdt} {et~al.}(2000){Breitschwerdt}, {Freyberg}, \& {Egger}}]{Breitschwerdt2000}
{Breitschwerdt}, D., {Freyberg}, M.~J., \& {Egger}, R. 2000, \aap, 361, 303

\bibitem[{{Castor} {et~al.}(1975){Castor}, {McCray}, \& {Weaver}}]{Castor1975}
{Castor}, J., {McCray}, R., \& {Weaver}, R. 1975, \apjl, 200, L107

\bibitem[{{Chang} \& {Kiang}(2024)}]{Chang2024}
{Chang}, C.-J. \& {Kiang}, J.-F. 2024, Universe, 10, 279

\bibitem[{{Chatzikos} {et~al.}(2023){Chatzikos}, {Bianchi}, {Camilloni}, {Chakraborty}, {Gunasekera}, {Guzm{\'a}n}, {Milby}, {Sarkar}, {Shaw}, {van Hoof}, \& {Ferland}}]{Chatzikos2023}
{Chatzikos}, M., {Bianchi}, S., {Camilloni}, F., {et~al.} 2023, \rmxaa, 59, 327

\bibitem[{{Cheng} {et~al.}(2011){Cheng}, {Chernyshov}, {Dogiel}, {Ko}, \& {Ip}}]{Cheng2011}
{Cheng}, K.~S., {Chernyshov}, D.~O., {Dogiel}, V.~A., {Ko}, C.~M., \& {Ip}, W.~H. 2011, \apjl, 731, L17

\bibitem[{{Cordes} {et~al.}(1991){Cordes}, {Weisberg}, {Frail}, {Spangler}, \& {Ryan}}]{Cordes1991}
{Cordes}, J.~M., {Weisberg}, J.~M., {Frail}, D.~A., {Spangler}, S.~R., \& {Ryan}, M. 1991, \nat, 354, 121

\bibitem[{{Crocker} {et~al.}(2015){Crocker}, {Bicknell}, {Taylor}, \& {Carretti}}]{Crocker2015}
{Crocker}, R.~M., {Bicknell}, G.~V., {Taylor}, A.~M., \& {Carretti}, E. 2015, \apj, 808, 107

\bibitem[{{Crocker} {et~al.}(2011){Crocker}, {Jones}, {Aharonian}, {et~al.}}]{Crocker2011}
{Crocker}, R.~M., {Jones}, D.~I., {Aharonian}, F., {et~al.} 2011, \mnras, 413, 763

\bibitem[{Dalcin {et~al.}(2011)Dalcin, Paz, Kler, \& Cosimo}]{Dalcin2011}
Dalcin, L.~D., Paz, R.~R., Kler, P.~A., \& Cosimo, A. 2011, Advances in Water Resources, 34, 1124, new Computational Methods and Software Tools

\bibitem[{{Dauser} {et~al.}(2019){Dauser}, {Falkner}, {Lorenz}, {Kirsch}, {Peille}, {Cucchetti}, {Schmid}, {Brand}, {Oertel}, {Smith}, \& {Wilms}}]{Dauser2019}
{Dauser}, T., {Falkner}, S., {Lorenz}, M., {et~al.} 2019, \aap, 630, A66

\bibitem[{{Dorfi} \& {Breitschwerdt}(2012)}]{DB:12}
{Dorfi}, E.~A. \& {Breitschwerdt}, D. 2012, \aap, 540, A77

\bibitem[{{Dutta} {et~al.}(2024){Dutta}, {Sharma}, {Sarkar}, \& {Stone}}]{Dutta2024}
{Dutta}, R., {Sharma}, P., {Sarkar}, K.~C., \& {Stone}, J.~M. 2024, \apj, 973, 148

\bibitem[{{Event Horizon Telescope Collaboration} {et~al.}(2022){Event Horizon Telescope Collaboration}, {Akiyama}, {Alberdi}, {et~al.}}]{EHT2022}
{Event Horizon Telescope Collaboration}, {Akiyama}, K., {Alberdi}, A., {et~al.} 2022, \apjl, 930, L12

\bibitem[{{Feldman}(1992)}]{Feldman1992}
{Feldman}, U. 1992, Physica Scripta Volume T, 46, 202

\bibitem[{{Ferri{\`e}re}(2009)}]{Ferriere2009}
{Ferri{\`e}re}, K. 2009, \aap, 505, 1183

\bibitem[{{Guo} \& {Mathews}(2012)}]{Guo2012}
{Guo}, F. \& {Mathews}, W.~G. 2012, \apj, 756, 181

\bibitem[{Górski {et~al.}(2005)Górski, Hivon, Banday, Wandelt, Hansen, Reinecke, \& Bartelmann}]{Gorski2005}
Górski, K.~M., Hivon, E., Banday, A.~J., {et~al.} 2005, \apj, 622, 759

\bibitem[{Harris {et~al.}(2020)Harris, Millman, van~der Walt, Gommers, Virtanen, Cournapeau, Wieser, Taylor, Berg, Smith, Kern, Picus, Hoyer, van Kerkwijk, Brett, Haldane, del R{\'{i}}o, Wiebe, Peterson, G{\'{e}}rard-Marchant, Sheppard, Reddy, Weckesser, Abbasi, Gohlke, \& Oliphant}]{Harris2020}
Harris, C.~R., Millman, K.~J., van~der Walt, S.~J., {et~al.} 2020, Nature, 585, 357

\bibitem[{{HI4PI Collaboration} {et~al.}(2016){HI4PI Collaboration}, {Ben Bekhti}, {Fl{\"o}er}, {Keller}, {Kerp}, {Lenz}, {Winkel}, {Bailin}, {Calabretta}, {Dedes}, {Ford}, {Gibson}, {Haud}, {Janowiecki}, {Kalberla}, {Lockman}, {McClure-Griffiths}, {Murphy}, {Nakanishi}, {Pisano}, \& {Staveley-Smith}}]{HI4PI2016}
{HI4PI Collaboration}, {Ben Bekhti}, N., {Fl{\"o}er}, L., {et~al.} 2016, \aap, 594, A116

\bibitem[{Hunter(2007)}]{Hunter2007}
Hunter, J.~D. 2007, Computing in Science \& Engineering, 9, 90

\bibitem[{{Jokipii}(1982)}]{Jokipii1982}
{Jokipii}, J.~R. 1982, \apj, 255, 716

\bibitem[{{Khabibullin} \& {Churazov}(2019)}]{Khabibullin2019}
{Khabibullin}, I. \& {Churazov}, E. 2019, \mnras, 482, 4972

\bibitem[{{Ko} {et~al.}(2020){Ko}, {Breitschwerdt}, {Chernyshov}, {et~al.}}]{Ko2020}
{Ko}, C.~M., {Breitschwerdt}, D., {Chernyshov}, {et~al.} 2020, \apj, 904, 46

\bibitem[{{Kompaneets}(1960)}]{Kompaneets1960}
{Kompaneets}, D.~A. 1960, Akademiia Nauk SSSR Doklady, 130, 1001

\bibitem[{{Lacy} {et~al.}(1982){Lacy}, {Townes}, \& {Hollenbach}}]{Lacy1982}
{Lacy}, J.~H., {Townes}, C.~H., \& {Hollenbach}, D.~J. 1982, \apj, 262, 120

\bibitem[{{Liu} {et~al.}(2024){Liu}, {Merloni}, {Sanders}, {Ponti}, {Strong}, {Yeung}, {Locatelli}, {Predehl}, {Zheng}, {Sasaki}, {Freyberg}, {Dennerl}, {Becker}, {Nandra}, {Mayer}, \& {Buchner}}]{Liu2024}
{Liu}, T., {Merloni}, A., {Sanders}, J., {et~al.} 2024, \apjl, 967, L27

\bibitem[{{Magorrian} \& {Tremaine}(1999)}]{Magorrian1999}
{Magorrian}, J. \& {Tremaine}, S. 1999, \mnras, 309, 447

\bibitem[{{McMillan}(2017)}]{McMillan2017}
{McMillan}, P.~J. 2017, \mnras, 465, 76

\bibitem[{{Miller} \& {Bregman}(2013)}]{Miller2013}
{Miller}, M.~J. \& {Bregman}, J.~N. 2013, \apj, 770, 118

\bibitem[{{Mou} {et~al.}(2023){Mou}, {Sun}, {Fang}, {Wang}, {Zhang}, {Yuan}, {Sofue}, {Wang}, \& {He}}]{Mou2023}
{Mou}, G., {Sun}, D., {Fang}, T., {et~al.} 2023, Nature Communications, 14, 781

\bibitem[{{Nakashima} {et~al.}(2018){Nakashima}, {Inoue}, {Yamasaki}, {et~al.}}]{Nakashima2018}
{Nakashima}, S., {Inoue}, Y., {Yamasaki}, N., {et~al.} 2018, \apj, 862, 34

\bibitem[{{O'Sullivan} {et~al.}(2003){O'Sullivan}, {Ponman}, \& {Collins}}]{Osullivan2003}
{O'Sullivan}, E., {Ponman}, T.~J., \& {Collins}, R.~S. 2003, \mnras, 340, 1375

\bibitem[{{Patankar}(1980)}]{Patankar1980}
{Patankar}, S.~V. 1980, {Numerical heat transfer and fluid flow} (CRC press)

\bibitem[{{Pfrommer} {et~al.}(2017){Pfrommer}, {Pakmor}, {Schaal}, {Simpson}, \& {Springel}}]{Pfrommer2017}
{Pfrommer}, C., {Pakmor}, R., {Schaal}, K., {Simpson}, C.~M., \& {Springel}, V. 2017, \mnras, 465, 4500

\bibitem[{{Ponti} {et~al.}(2015){Ponti}, {Morris}, {Terrier}, {et~al.}}]{Ponti2015}
{Ponti}, G., {Morris}, M.~R., {Terrier}, R., {et~al.} 2015, \mnras, 453, 172

\bibitem[{{Predehl} {et~al.}(2021){Predehl}, {Andritschke}, {Arefiev}, {Babyshkin}, {Batanov}, {Becker}, {B{\"o}hringer}, {Bogomolov}, {Boller}, {Borm}, {Bornemann}, {Br{\"a}uninger}, {Br{\"u}ggen}, {Brunner}, {Brusa}, {Bulbul}, {Buntov}, {Burwitz}, {Burkert}, {Clerc}, {Churazov}, {Coutinho}, {Dauser}, {Dennerl}, {Doroshenko}, {Eder}, {Emberger}, {Eraerds}, {Finoguenov}, {Freyberg}, {Friedrich}, {Friedrich}, {F{\"u}rmetz}, {Georgakakis}, {Gilfanov}, {Granato}, {Grossberger}, {Gueguen}, {Gureev}, {Haberl}, {H{\"a}lker}, {Hartner}, {Hasinger}, {Huber}, {Ji}, {Kienlin}, {Kink}, {Korotkov}, {Kreykenbohm}, {Lamer}, {Lomakin}, {Lapshov}, {Liu}, {Maitra}, {Meidinger}, {Menz}, {Merloni}, {Mernik}, {Mican}, {Mohr}, {M{\"u}ller}, {Nandra}, {Nazarov}, {Pacaud}, {Pavlinsky}, {Perinati}, {Pfeffermann}, {Pietschner}, {Ramos-Ceja}, {Rau}, {Reiffers}, {Reiprich}, {Robrade}, {Salvato}, {Sanders}, {Santangelo}, {Sasaki}, {Scheuerle}, {Schmid}, {Schmitt}, {Schwope}, {Shirshakov}, {Steinmetz}, {Stewart}, {Str{\"u}der},
  {Sunyaev}, {Tenzer}, {Tiedemann}, {Tr{\"u}mper}, {Voron}, {Weber}, {Wilms}, \& {Yaroshenko}}]{Predehl2021}
{Predehl}, P., {Andritschke}, R., {Arefiev}, V., {et~al.} 2021, \aap, 647, A1

\bibitem[{{Predehl} {et~al.}(2020){Predehl}, {Sunyaev}, {Becker}, {et~al.}}]{Predehl2020}
{Predehl}, P., {Sunyaev}, R.~A., {Becker}, W., {et~al.} 2020, \nat, 588, 227

\bibitem[{{Rayleigh}(1882)}]{Rayleigh1882}
{Rayleigh}, L. 1882, Proceedings of the London Mathematical Society, s1-14, 170

\bibitem[{{Rieke} {et~al.}(1980){Rieke}, {Lebofsky}, {Thompson}, {Low}, \& {Tokunaga}}]{Rieke1980}
{Rieke}, G.~H., {Lebofsky}, M.~J., {Thompson}, R.~I., {Low}, F.~J., \& {Tokunaga}, A.~T. 1980, \apj, 238, 24

\bibitem[{{Sarkar}(2019)}]{Sarkar2019}
{Sarkar}, K.~C. 2019, \mnras, 482, 4813

\bibitem[{{Sarkar}(2024)}]{Sarkar2024}
{Sarkar}, K.~C. 2024, \aapr, 32, 1

\bibitem[{{Sarkar} {et~al.}(2023){Sarkar}, {Mondal}, {Sharma}, \& {Piran}}]{Sarkar2023}
{Sarkar}, K.~C., {Mondal}, S., {Sharma}, P., \& {Piran}, T. 2023, \apj, 951, 36

\bibitem[{{Schaal} \& {Springel}(2015)}]{Schaal2015}
{Schaal}, K. \& {Springel}, V. 2015, \mnras, 446, 3992

\bibitem[{{Schulreich} \& {Breitschwerdt}(2022)}]{Schulreich2022}
{Schulreich}, M.~M. \& {Breitschwerdt}, D. 2022, \mnras, 509, 716

\bibitem[{{Shao} {et~al.}(2011){Shao}, {Zhang}, {Fan}, \& {Wei}}]{Shao2011}
{Shao}, L., {Zhang}, F.-W., {Fan}, Y.-Z., \& {Wei}, D.-M. 2011, \apjl, 734, L33

\bibitem[{{Su} {et~al.}(2010){Su}, {Slatyer}, \& {Finkbeiner}}]{Su2010}
{Su}, M., {Slatyer}, T.~R., \& {Finkbeiner}, D.~P. 2010, \apj, 724, 1044

\bibitem[{{Taylor}(1950)}]{Taylor1950}
{Taylor}, G. 1950, Proceedings of the Royal Society of London Series A, 201, 192

\bibitem[{{Teyssier}(2002)}]{Teyssier2002}
{Teyssier}, R. 2002, \aap, 385, 337

\bibitem[{Toro(1992)}]{Toro1992}
Toro, E.~F. 1992, Philosophical Transactions: Physical Sciences and Engineering, 341, 499

\bibitem[{{Toro}(1997)}]{Toro1997}
{Toro}, E.~F. 1997, {\normalfont Riemann Solvers and Numerical Methods for Fluid Dynamics: A Practical Introduction} (\textit{Springer Berlin Heidelberg})

\bibitem[{{Toro} {et~al.}(1994){Toro}, {Spruce}, \& {Speares}}]{Toro1994}
{Toro}, E.~F., {Spruce}, M., \& {Speares}, W. 1994, Shock Waves, 4, 25

\bibitem[{{Tseng} {et~al.}(2024){Tseng}, {Yang}, {Chen}, {Schive}, \& {Chiueh}}]{Tseng2024}
{Tseng}, P.-H., {Yang}, H. Y.~K., {Chen}, C.-Y., {Schive}, H.-Y., \& {Chiueh}, T. 2024, \apj, 970, 146

\bibitem[{{Turk} {et~al.}(2011){Turk}, {Smith}, {Oishi}, {Skory}, {Skillman}, {Abel}, \& {Norman}}]{Turk2011}
{Turk}, M.~J., {Smith}, B.~D., {Oishi}, J.~S., {et~al.} 2011, The Astrophysical Journal Supplement Series, 192, 9

\bibitem[{{Weaver} {et~al.}(1977){Weaver}, {McCray}, {Castor}, {Shapiro}, \& {Moore}}]{Weaver1977}
{Weaver}, R., {McCray}, R., {Castor}, J., {Shapiro}, P., \& {Moore}, R. 1977, \apj, 218, 377

\bibitem[{{Wilms} {et~al.}(2000){Wilms}, {Allen}, \& {McCray}}]{Wilms2000}
{Wilms}, J., {Allen}, A., \& {McCray}, R. 2000, \apj, 542, 914

\bibitem[{{Yang} {et~al.}(2012){Yang}, {Ruszkowski}, {Ricker}, {Zweibel}, \& {Lee}}]{Yang2012}
{Yang}, H. Y.~K., {Ruszkowski}, M., {Ricker}, P.~M., {Zweibel}, E., \& {Lee}, D. 2012, \apj, 761, 185

\bibitem[{{Yang} {et~al.}(2022){Yang}, {Ruszkowski}, \& {Zweibel}}]{Yang2022}
{Yang}, H. Y.~K., {Ruszkowski}, M., \& {Zweibel}, E.~G. 2022, Nature Astronomy, 6, 584

\bibitem[{{Zhang} \& {Guo}(2020)}]{Zhang2020}
{Zhang}, R. \& {Guo}, F. 2020, \apj, 894, 117

\bibitem[{Zonca {et~al.}(2019)Zonca, Singer, Lenz, Reinecke, Rosset, Hivon, \& Gorski}]{Zonca2019}
Zonca, A., Singer, L., Lenz, D., {et~al.} 2019, Journal of Open Source Software, 4, 1298

\bibitem[{{Z}uHone {et~al.}(2014){Z}uHone, {B}iffi, {H}allman, {R}andall, {F}oster, \& {S}chmid}]{ZuHone2014}
{Z}uHone, J.~A., {B}iffi, V., {H}allman, E.~J., {et~al.} 2014, in {P}roceedings of the 13th {P}ython in {S}cience {C}onference, ed. {S}t\'efan van~der {W}alt \& {J}ames {B}ergstra, 98 -- 104

\bibitem[{{ZuHone} {et~al.}(2023){ZuHone}, {Vikhlinin}, {Tremblay}, {Randall}, {Andrade-Santos}, \& {Bourdin}}]{ZuHone2023}
{ZuHone}, J.~A., {Vikhlinin}, A., {Tremblay}, G.~R., {et~al.} 2023, {SOXS: Simulated Observations of X-ray Sources}, Astrophysics Source Code Library, record ascl:2301.024

\end{thebibliography}
\end{document}